\documentclass[11pt,a4paper]{article}
\pdfoutput=1
\usepackage{jheppub}
\usepackage{epsfig,simplewick,xcolor,amsmath}%,ytableau}
\usepackage{fancybox}
\usepackage[vcentermath]{youngtab}

\def\sym{\rm sym}

\def\sgn{\rm sgn}

\def \Tr{\mbox{Tr\,}}

\newcommand{\be}{\begin{equation}}
\newcommand{\bea}{\begin{eqnarray}}
\newcommand{\ee}{\end{equation}}
\newcommand{\eea}{\end{eqnarray}}

\begin{document}
\title{Structure of Loop Space at Finite $N$}

\author[a]{Robert de Mello Koch}
\affiliation[a]{School of Science, Huzhou University, Huzhou 313000, China}
\affiliation[a]{Mandelstam Institute for Theoretical Physics, School of Physics, University of the Witwatersrand, Private Bag 3, Wits 2050, South Africa}

\author[b]{and Antal Jevicki}
\affiliation[b]{Department of Physics, Brown University,
182 Hope Street, Providence, RI 02912, United States}
\affiliation[b]{Brown Theoretical Physics Center, Brown University,
340 Brook Street, Providence, RI 02912, United States}

\abstract{The space of invariants for a single matrix is generated by traces containing at most $N$ matrices per trace. We extend this analysis to multi-matrix models at finite $N$. Using the Molien-Weyl formula, we compute partition functions for various multi-matrix models at different $N$ and interpret them through trace relations. This allows us to identify a complete set of invariants, naturally divided into two distinct classes: primary and secondary. The full invariant ring of the multi-matrix model is reconstructed via the Hironaka decomposition, where primary invariants act freely, while secondary invariants satisfy quadratic relations. Significantly, while traces with at most $N$ matrices are always present, we also find invariants involving more than $N$ matrices per trace. The primary invariants correspond to perturbative degrees of freedom, whereas the secondary invariants emerge as non-trivial background structures. The growth of secondary invariants aligns with expectations from black hole entropy, suggesting deep structural connections to gravitational systems.}

\maketitle

{\vskip -2.5cm}

\section{Introduction}

In this paper, we investigate the space of gauge-invariant operators in the matrix quantum mechanics of $d$ Hermitian matrices $X^a$, with $a = 1, \dots, d$, transforming under the adjoint action of the unitary group:
\begin{align}
X^a \to U^\dagger X^a U.
\end{align}
The gauge-invariant operators -- which are the collective fields in a collective field theory description \cite{Jevicki:1979mb,deMelloKoch:2002nq} -- are polynomials in traces of words constructed from the matrices $X^a$. Our goal is to determine the complete ring of such invariants (loop space) at finite $N$, and to understand their algebraic structure. 

This problem reduces to identifying the trace relations and determining a minimal, independent generating set. These trace relations appear as constraints that commute with the collective field theory Hamiltonian. The eigenstates of the free matrix model are given by Schur polynomials \cite{Jevicki:1991yi} in the single-matrix case, and by restricted Schur polynomials \cite{Bhattacharyya:2008rb,Bhattacharyya:2008xy} in the multi-matrix setting. In this basis, trace relations manifest as null states -- states with zero or negative energy eigenvalues. Thus, the complete set of trace relations is captured by Schur or restricted Schur polynomials labeled by Young diagrams with more than $N$ rows. For further useful background and for other bases using Young diagram methods see \cite{Corley:2001zk,Kimura:2007wy,Brown:2007xh,Brown:2008ij}.

To illustrate the problem in the simplest setting, consider a single matrix $X$. Since $X$ has only $N$ independent eigenvalues, not all traces $\phi_n = \Tr(X^n)$ are independent. In fact, the ring of gauge-invariant operators is generated by $\phi_n$ with $n \leq N$, while for $n > N$, $\phi_n$ can be expressed as a polynomial in the lower-degree traces via trace relations\cite{Jevicki:2024fnk}. This finite truncation is a hallmark of finite-$N$ effects. In this work, we extend this analysis to multi-matrix models, providing a complete and constructive understanding of the ring of invariants at finite $N$.

Finite-$N$ effects have also played a central role in recent studies of black hole microstates, particularly in the discovery of the fortuity mechanism \cite{Chang:2022mjp,Choi:2022caq,Choi:2023znd,Chang:2023zqk,Choi:2023vdm,Chang:2024zqi}. These microstates are $\frac{1}{16}$-BPS and fall into two classes: monotone and fortuitous. Monotone states remain BPS at all $N$, while fortuitous states lose their BPS nature above a critical $N$. In the latter case, BPS saturation depends crucially on trace relations. For related work in the SYK model and the D1-D5 system, see \cite{Chang:2024lxt,Chang:2025rqy}.

Similarly  one is interested in the role of  finite-$N$ effects for thermal properties\cite{Sundborg:1999ue,Aharony:2003sx}  of multi-matrix models. In the case of $O(N)$ vector models  \cite{Shenker:2011zf} it was seen that they become significant at temperatures $T \sim \sqrt{N}$. The basic gauge-invariant operators -- bilocals -- cease to be independent once their number exceeds $N$, necessitating a cutoff and leading to modified entropy scaling $S \sim N T^2 V$.

We  examine the appearance of trace relations (at finite, integer N) from three complementary perspectives in Section~\ref{TraceRel}. First, trace relations correspond to states annihilated by the matrix Laplacian \cite{Jevicki:1991yi}. Second, invariant theory \cite{sneddon,Procesi} tells us that all relations follow from antisymmetrizing $N+1$ indices, each taking $N$ values. Third, Procesi \cite{Procesi} has shown that the full set of trace relations follows from the Cayley-Hamilton theorem. These perspectives enable both explicit construction and enumeration of trace relations.

In Section~\ref{partitionfuncs}, we discuss the decomposition of finite-$N$ partition functions using the Molien-Weyl formula. For $N=2$, we analyze models with two, three, and four matrices; for $N=3$, we examine two- and three-matrix models. For $N=4$ through $7$, we restrict attention to the two-matrix case due to increasing complexity. These partition functions take the form
\bea
Z(x) = \frac{1 + \sum_i c^s_i x^i}{\prod_j (1 - x^j)^{c^m_j}}. \label{illpf}
\eea
This is the Hilbert series of the invariant ring $C_{N,d}$ of $GL(N)$ invariants, and it matches the structure predicted by the Hironaka decomposition \cite{sturmfels,Grinstein:2023njq}. The denominator encodes \emph{primary} invariants, while the numerator encodes \emph{secondary} invariants. The number of primary invariants equals the number of denominator factors and gives the Krull dimension of the ring: $(d-1)N^2 + 1$ \cite{Herstein}. The number of secondary invariants is the value of the numerator at $x=1$, minus one. Here, $c^s_i$ counts secondary invariants built from $i$ fields, and $c^m_j$ counts primary invariants built from $j$ fields.

That our partition functions all take this Hironaka form is key to our analysis. The Hochster-Roberts theorem \cite{HR} ensures that $C_{N,d}$ is Cohen-Macaulay, since $GL(N)$ is a linearly reductive group over a field of characteristic zero. Therefore, the ring admits a Hironaka decomposition, i.e., it is a free module over a polynomial subalgebra.

From the form of $Z(x)$ in \eqref{illpf}, it follows that when constructing a gauge invariant operator from the invariants, primary invariants may appear with arbitrary powers but secondary invariants appear at most linearly. This constraint arises directly from the trace relations. For low $N$, one can explicitly match traces to primary and secondary generators, and verify that all operators are polynomial functions of these generators. We carry this out fully for $N=2$ in the two- and three-matrix models, and for $N=3$ in the two-matrix case. Secondary invariants provide examples of \emph{quadratically reducible} operators: their quadratic products do not vanish but reduce to combinations of primary and secondary invariants, at most linear in the secondary invariants.

The two- and three-matrix models at $N=2$ are especially simple and admit a grading by matrix type. In contrast, for the four-matrix model at $N=2$, the graded partition function includes negative terms in the numerator, indicating relations between the generators. These correspond to constraints between the invariants, showing the ring is not freely generated under the finer matrix-type grading. We have identified the full set of such constraints. This does not invalidate the Hironaka decomposition; rather, it reflects the fact that this decomposition does not respect the refined grading. As discussed in Section~\ref{fourmatrixdiscusion}, the generators themselves cannot be consistently assigned a grading \cite{Teranishi39,Domokos}.

Trace relations provide a solid foundation for interpreting the partition function. For small $N$, we validate this through explicit computation. These examples establish a clear pattern that we extrapolate to larger $N$, where direct computation becomes impractical. Relying on the partition function, we observe that the number of primary invariants remains $(d-1)N^2 + 1$, while the number of secondary invariants grows rapidly, as $\sim e^{{\rm const} \times N^2}$.

Throughout the paper, we make extensive use of tools from the study of graded rings, especially Hilbert series and the Molien-Weyl formula \cite{sturmfels,bruns}. These methods are increasingly relevant in quantum field theory for enumerating invariants. For related applications, see \cite{Benvenuti:2006qr,Feng:2007ur,Gray:2008yu,Jenkins:2009dy,Hanany:2010vu,Lehman:2015via,Henning:2015daa,Lehman:2015coa,deMelloKoch:2017caf,deMelloKoch:2017dgi,Henning:2017fpj,Bijnens:2022zqo,Graf:2022rco,Sun:2022aag,Kobach:2017xkw,Grinstein:2023njq}.

Further discussion and implications of our results are presented in Section~\ref{discuss}.

\section{Trace Relations}\label{TraceRel}

Trace relations play a central role in our analysis. In this section, we present three complementary perspectives on their origin: (1) as states annihilated by the matrix Laplacian, (2) as consequences of antisymmetrizing $N+1$ indices, each taking $N$ values, and (3) as direct consequences of the Cayley-Hamilton theorem. Each viewpoint offers valuable insight into the structure and implications of the trace relations.

\subsection{Matrix Laplacian}

The Laplacian on a single $N\times N$ matrix is closely related to the Laplacian on $U(N)$ \cite{Jevicki:1991yi}. Its eigenfunctions are Schur polynomials, which correspond to characters of $U(N)$. For multiple matrices, the eigenfunctions generalize to restricted Schur polynomials.

The matrix Laplacian is defined as
\begin{align}
    H = -\frac{1}{2} \Tr \left( \frac{\partial}{\partial M} \frac{\partial}{\partial M} \right) = -\frac{1}{2} \sum_{i,j=1}^{N} \frac{\partial}{\partial M^i_j} \frac{\partial}{\partial M^j_i}.
\end{align}

Under the variable change $U = e^{iM}$, the Laplacian becomes
\begin{align}
    H = N \Tr \left( U \frac{\partial}{\partial U} \right) + : \Tr \left( U \frac{\partial}{\partial U} U \frac{\partial}{\partial U} \right) :
\end{align}
with normal ordering to prevent derivatives acting within the trace. The eigenfunctions are Schur polynomials \cite{Jevicki:1991yi}
\begin{align}
    \chi_R(U) = \frac{1}{n!} \sum_{\sigma \in S_n} \chi_R(\sigma) \Tr(\sigma U^{\otimes n}),
\end{align}
where $R$ is a Young diagram with $n$ boxes. The corresponding eigenvalue is\footnote{This result follows immediately from the identity
\bea
:\Tr\left(U{\partial\over\partial U}U{\partial\over\partial U}\right):U^{j_1}_{i_1}\cdots U^{j_n}_{i_n}&=&\sum_{\substack{a,b=1\\ a\ne b}}^N
U^{j_1}_{i_1}\cdots U^{j_a}_{i_b}\cdots U^{j_b}_{i_a}\cdots U^{j_n}_{i_n} 
\eea}
\begin{align}
    H \chi_R(U) = (nN + 2\lambda_2^R) \chi_R(U),
\end{align}
where the symmetric group Casimir is
\begin{align}
    \lambda_2^R = \sum_i \frac{r_i (r_i -1)}{2} - \sum_j \frac{c_j (c_j - 1)}{2},
\end{align}
with $r_i$ and $c_j$ the row and column lengths of $R$. Table~\ref{table:schur} shows examples.

\begin{table}[h!]
\centering
\renewcommand{\arraystretch}{1.5}
\begin{tabular}{|c|c|c|}
\hline
$R$ & $\lambda_2^R$ & $E$ \\
\hline\hline
{\tiny\yng(1)} & 0 & $N$ \\
\hline
{\tiny\yng(2)} & 1 & $2N+2$ \\
\hline
{\tiny\yng(1,1)} & -1 & $2N-2$ \\
\hline
{\tiny\yng(3)} & 3 & $3N+6$ \\
\hline
{\tiny\yng(2,1)} & 0 & $3N$ \\
\hline
{\tiny\yng(1,1,1)} & -3 & $3N-6$ \\
\hline
\end{tabular}
\caption{Young diagrams, Casimir values, and energy eigenvalues.}
\label{table:schur}
\end{table}

For $N=2$, $\chi_{\tiny\yng(1,1,1)}$ is both a zero eigenvalue and a null state. More generally, Schur polynomials labeled by Young diagrams with more than $N$ rows are null states. To define the problem consistently, we first solve for general $N$ and then restrict to finite $N$.

Schur polynomials with at most $N$ rows form a basis for single-matrix invariants. Those with more than $N$ rows generate all trace relations.

This extends naturally to multi-matrix models. For two matrices $X$ and $Y$, the Laplacian becomes
\begin{align}
    H = -\frac{1}{2} \sum_{i,j=1}^{N} \left( \frac{\partial^2}{\partial X^i_j \partial X^j_i} + \frac{\partial^2}{\partial Y^i_j \partial Y^j_i} \right).
\end{align}
Changing variables to $U_X = e^{iX}$ and $U_Y = e^{iY}$, we find
\begin{align}
    H = \sum_{a=X,Y} \left( N \Tr \left( U_a \frac{\partial}{\partial U_a} \right) + : \Tr \left( U_a \frac{\partial}{\partial U_a} U_a \frac{\partial}{\partial U_a} \right) : \right).
\end{align}
The eigenfunctions are restricted Schur polynomials \cite{Bhattacharyya:2008rb,Bhattacharyya:2008xy}
\begin{align}
\chi_{R,(r,s)\alpha\beta}(U_X,U_Y) = \frac{1}{n!m!} \sum_{\sigma \in S_{n+m}} \chi_{R,(r,s)\alpha\beta}(\sigma) \Tr(\sigma U_X^{\otimes n} U_Y^{\otimes m}),
\end{align}
$\chi_{R,(r,s)\alpha\beta}(\sigma)$ is a restricted character in the terminology of \cite{deMelloKoch:2007rqf}. The character is a trace over the matrix representing $\sigma$ in irreducible representation $R$. The restricted character restricts the trace to a representation $(r,s)$ of the $S_n\times S_m$ subgroup. After restricting to the subgroup, $f_{rsR}$ copies of the $(r,s)$ representation are obtained. Here $f_{rsR}$ is the usual Littlewood-Richardson coefficient. The multiplicity index $\alpha$ specifies which copy is used for the row index in the trace while $\beta$ specifies which copy of the column index is used. The $S_n\times S_m$ group permutes indices of $U_X$ with each other and, independently, $U_Y$ with each other. A completely parallel argument to the one given above shows that
\bea
H\chi_{R,(r,s)\alpha\beta}(U_X,U_Y)&=&(nN+2\lambda^r_2+2\lambda_2^s)\chi_{R,(r,s)\alpha\beta}(U_X,U_Y)
\eea
Trace relations correspond to eigenstates with vanishing or negative energy and again, these eigenstates are actually null states. The restricted Schur polynomials $\chi_{R,(r,s)\alpha\beta}(U_X,U_Y)$ labelled with Young diagrams $R$ with no more than $N$ rows provide a basis for two matrix wave functions, while the restricted Schur polynomials labelled by Young diagrams $R$ with more than $N$ rows provide a complete set of trace relations for two matrices.

The Schur polynomials for a general multi-matrix model with $d$ species of matrix $X^a$ are labelled by $d+1$ Young diagrams, one diagram for each species and one which organizes indices of all fields. These are eigenstates of the Hamiltonian
\bea
    H = \sum_{a=1,2,\cdots,d} \left( N \Tr \left( U_a \frac{\partial}{\partial U_a} \right) + : \Tr \left( U_a \frac{\partial}{\partial U_a} U_a \frac{\partial}{\partial U_a} \right) : \right).
\eea
where $U_a=e^{iX^a}$. Trace relations again correspond to states with vanishing or negative energy. Restricted Schur polynomials labelled with Young diagrams with no more than $N$ rows provide a basis for the multi-matrix wave functions, while those labelled by Young diagrams with more than $N$ rows provide a complete set of trace relations.

\subsection{Invariant Theory}\label{invth}

An invariant is a polynomial function of tensors that remains unchanged under group transformations. For matrices $(X^a)^i_j$ transforming under $U(N)$ as $X^a \to UX^aU^\dagger$, we are interested in two questions: what are the invariants, and what are the relations among them \cite{sneddon,VD,sturmfels}.

{\vskip 0.1cm}

\textbf{First Fundamental Theorem:} Any polynomial invariant can be expressed as a linear combination of complete contractions of products of tensors.

{\vskip 0.1cm}

\textbf{Second Fundamental Theorem:} All identities among invariants follow from antisymmetrizing over $N+1$ indices in $N$ dimensions.

{\vskip 0.1cm}

The first theorem ensures our invariants are multi-trace operators. The second implies that all trace relations come from Schur (or restricted Schur) polynomials labeled by diagrams with more than $N$ rows, since columns correspond to antisymmetrization. This connection allows counting of independent gauge-invariant operators and trace relations (see Appendix~\ref{CountSandRS}). Other bases for invariants labeled by Young diagrams also exist \cite{Kimura:2007wy,Brown:2007xh,Brown:2008ij} and can be used for this analysis.

Lastly, some identities apply to the matrices themselves, not just their traces. The Amitsur--Levitzki theorem (see for example \cite{VD}) states that the ring of $n \times n$ matrices satisfies a standard polynomial identity of degree $2n$, which is the minimal such identity. Substituting any $2n$ matrices into the standard alternating polynomial yields zero.

\subsection{Cayley-Hamilton Theorem}

Procesi \cite{Procesi} proved that the ring of invariants of $d$ $N \times N$ matrices is finitely generated by traces of words of length $\leq 2^N - 1$ (Theorem 3.4a), and that all relations among these invariants follow from the Cayley-Hamilton theorem (Theorem 4.6).

This is extremely useful for identifying trace relations: they can always be expressed as
\begin{align}
0 = \sum_{\sigma \in S_{N+1}} \text{sgn}(\sigma) (A_1)^{i_1}_{i_{\sigma(1)}} (A_2)^{i_2}_{i_{\sigma(2)}} \cdots (A_{N+1})^{i_{N+1}}_{i_{\sigma(N+1)}},
\end{align}
where each $A_i$ is a word in the matrices $X^a$, and $\text{sgn}(\sigma)$ is the sign of the permutation. All trace relations are consequences of this identity.

\section{Decomposition}\label{partitionfuncs}

In this section the correspondence between loop space decomposition and decomposition of the finite partition function is established. The main tool is the Molien-Weyl formula \cite{MW1,MW2,MW3} which computes the partition function for fixed values of $N$ in various free matrix models. The Molien-Weyl formula, whose derivation is reviewed in Appendix \ref{MolienWeylPartitionFunctions}, is efficiently evaluated using residue techniques. The resulting structure of the partition function is governed by the fact that the space of gauge-invariant operators admits a Hironaka decomposition. Before proceeding to explicit computations, we review this decomposition in the next subsection. We then turn to the evaluation of partition functions. For $N=2$, we analyze two-, three-, and four-matrix models, while for $N=3$, we consider two- and three-matrix models. For $N=4,5,6,7$, our focus is on two-matrix models. The resulting partition functions have an interpretation in terms of generators of two types: those that act freely (primary invariants) and those that are quadratically reducible (secondary invariants). For $N=2$ and $N=3$, we confirm this interpretation explicitly by utilizing trace relations.

\subsection{Hironaka Decomposition}

The partition functions computed in this work take the form (\ref{illpf}), as stated in the introduction. This structure arises because the complete space of gauge-invariant operators admits a Hironaka decomposition. This decomposition involves two distinct classes of invariants:

\begin{itemize}
\item \textbf{Primary invariants} \( \{m_i\} \) are algebraically independent.
\item \textbf{Secondary invariants} \( \{s_i\} \) satisfy relations of the form
\begin{align}
    s_k s_m = \sum_j f^j_{km} s_j,
\end{align}
where the coefficients \( f^j_{km} \) belong to the ring \( K[m_1,\dots,m_l] \) generated by the primary invariants, and we define \( s_0 = 1 \).
\end{itemize}

The full ring of gauge-invariant operators is then given by
\begin{align}
    \sum_j s_j K[m_1,\dots,m_l].
\end{align}

The Hilbert series encodes the graded count of these invariants. Thanks to the Hironaka decomposition, the Hilbert series can be written immediately as a rational function. The denominator encodes the primary invariants, while the numerator captures the secondary invariants. We grade the counting by the number of matrices in the trace.
The denominator is
\begin{align}
    D(x) = \prod_{i=1}^{d_P}(1 - x^{n_i}),
\end{align}
where \( d_P \) is the number of primary invariants, and \( n_i \) is the degree (number of matrices) of the \( i \)th primary invariant.
The numerator is
\begin{align}
    N(x) = 1 + \sum_{j=1}^{d_S} x^{t_j},
\end{align}
where \( d_S \) is the number of secondary invariants and \( t_j \) is the degree of the \( j \)th secondary invariant. We interpret the partition functions computed below as Hilbert series associated with a Hironaka decomposition.

As a simple example, the gauge-invariant operators of the one-matrix model are generated by the traces \( \phi_n = \Tr(X^n) \) for \( n = 1,2,\dots,N \). These \( \phi_n \) are algebraically independent, forming the set of primary invariants, with no secondary invariants. The Hilbert series, which coincides with the partition function of the free one-matrix model, is
\begin{align}
    Z(x) = \frac{1}{(1 - x)(1 - x^2)\cdots(1 - x^N)}.
\end{align}

Importantly, the set of primary and secondary invariants does not, in general, coincide with a minimal generating set for the ring. To illustrate this, consider a ring generated by a set of algebraically independent invariants \( \{m_i\} \) and a single element \( S \) satisfying
\begin{align}
    S^3 + c_1 S^2 + c_2 S + c_3 = 0,
\end{align}
with coefficients \( c_i \in K[m_1,\dots,m_l] \). This ring has \( l + 1 \) generators.
Now define secondary invariants
\begin{align}
    s_1 = S, \quad s_2 = S^2,
\end{align}
with \( s_0 = 1 \). These satisfy
\begin{align}
    s_1^2 &= s_2, \\
    s_1 s_2 &= -c_1 s_2 - c_2 s_1 - c_3, \\
    s_2^2 &= c_1(c_1 s_2 + c_2 s_1 + c_3) - c_2 s_2 - c_3 s_1,
\end{align}
showing that any quadratic product of secondary invariants reduces to a combination of \( s_0, s_1, s_2 \). Thus, although the ring has \( l+1 \) generators, we have \( l+2 \) primary and secondary invariants (excluding \( s_0 \)).

More generally, if a generator \( S \) satisfies
\begin{align}
    S^k + c_1 S^{k-1} + c_2 S^{k-2} + \cdots + c_k = 0,
\end{align}
with \( c_i \in K[m_1,\dots,m_l] \), then \( S \) is replaced by \( k-1 \) secondary invariants \( s_j = S^j \), for \( j = 1,2,\dots,k-1 \). These are all generated by \( s_1 \), and the number of primary and secondary invariants is always greater than or equal to the number of ring generators.

\subsection{$N=2$}

At $N=2$, the analysis remains relatively tractable, allowing us to explore the two-, three-, and four-matrix models in detail. The two-matrix model is particularly simple: its complete set of gauge-invariant operators is freely generated by five primary invariants. Using trace relations, we explicitly demonstrate that these five invariants are sufficient to generate the entire ring of invariants.

The three-matrix model introduces a new feature. In addition to nine primary invariants, it includes a single secondary invariant. We show, via trace relations, that this secondary invariant cannot be constructed solely from the primary invariants. Furthermore, we derive a constraint indicating that the square of the secondary invariant can be written as a polynomial in both primary and secondary invariants, where the secondary invariant appears at most linearly. This establishes that the secondary invariant is quadratically reducible.

The four-matrix model exhibits the richest structure. When the partition function is graded by matrix species, it reveals the presence of constraints among the generators. A detailed analysis using trace relations confirms this structure and reinforces the interpretation of the partition function, revealing a nontrivial interplay between the generators. The presence of constraints implies that the ring of invariants is not freely generated under the refined matrix-type grading. Upon relaxing this grading to consider only the total degree, the partition function admits a Hironaka decomposition. The construction of the associated primary and secondary invariants is already known in the literature. From their explicit form, it is evident that these invariants do not respect a grading by matrix species.

\subsubsection{Two matrix model}\label{2M2N}

Consider a free matrix model with two matrix species, $X$ and $Y$. To compute the exact finite-$N$ partition function, we apply the Molien-Weyl formula (\ref{MWPF}). Both matrices are assigned energy $E_X = E_Y = 1$. Introducing chemical potentials $\mu_x$ and $\mu_y$ to keep track of each matrix. In terms of the variables  
\bea
x=e^{-\beta E_1 - \mu_x}, \qquad y=e^{-\beta E_2 - \mu_y},
\eea
the Molien-Weyl formula (\ref{MWPF}) gives the partition function  
\bea
Z(x,y) = \left(\frac{1}{(1-x)(1-y)}\right)^2 \frac{1}{2\pi i} \oint_{\cal C} dt_1 \frac{t_1(1-t_1)}{(1-x t_1)(1-y t_1)(t_1-x)(t_1-y)},
\eea
where the integration contour ${\cal C}$ is the unit circle. For real chemical potentials and $\beta$, we have $x < 1$ and $y < 1$, so the integrand has poles inside the unit circle at $t = x$ and $t = y$. Evaluating the residues at these poles, we obtain  
\bea
Z(x,y) = \frac{1}{(1-x)(1-y)(1-x^2)(1-xy)(1-y^2)}.  
\label{exactN2M2}
\eea
Expanding the partition function as a power series produces terms with non-negative powers of $x$ and $y$. The coefficient of each monomial $x^n y^m$ is an integer counting the number of independent gauge-invariant operators constructed from $n$ $X$s and $m$ $Y$s. Using \texttt{Mathematica}, we have verified that the series expansion of (\ref{exactN2M2}) matches (\ref{LRPF}) exactly for all invariants involving up to 30 matrices. Beyond this point, evaluating the counting formula (\ref{LRPF}) becomes computationally expensive.

The partition function (\ref{exactN2M2}) suggests the complete space of gauge invariant operators is freely generated by all single trace operators with $\le N=2$ matrices in the trace
\bea
m_1 &=& \Tr(X), \qquad m_2\,\,=\,\, \Tr(Y), \cr\cr
m_3 &=& \Tr(X^2), \quad m_4\,\, =\,\, \Tr(XY), \quad m_5\,\, =\,\, \Tr(Y^2),
\label{mesonloopsN2M2}
\eea
which are all primary invariants. The proposed generating set (\ref{mesonloopsN2M2}) and partition function (\ref{exactN2M2}) were independently obtained in \cite{2mats} using graded ring theory techniques. There are no algebraic relations between free generators. Since for $N=2$ trace relations involve traces of three or more matrices, the fact that our generators are free is manifest.  

Invariants with more than two matrices per trace are generated from the primary invariants (\ref{mesonloopsN2M2}) via trace relations. The complete set of trace relations follows from the Cayley-Hamilton theorem, which for $N=2$, take the form $T_2(A,B,C) = 0$, where  
\bea
T_2(A,B,C) &= \Tr(A)\Tr(B)\Tr(C) - \Tr(AB)\Tr(C) - \Tr(AC)\Tr(B) \cr\cr
&\quad - \Tr(A)\Tr(BC) + \Tr(ABC) + \Tr(ACB),
\eea
and $A$, $B$ and $C$ are any words constructed using $X,Y$ as letters. Consider a single-trace invariant composed of $n+m$ letters, $n$ of which are $X$s and the remaining $m$ $Y$s. When $m+n=3$, each choice of $m$ and $n$ leads to a single trace relation, matching the number of possible gauge-invariant operators. For instance, choosing $m=2$ and $n=1$, we obtain  
\bea
T_2(X,Y,Y) = \Tr(X) \Tr(Y)^2 - 2 \Tr(XY) \Tr(Y) - \Tr(X) \Tr(XY) + 2 \Tr(XY^2) = 0,
\nonumber
\eea
which implies  
\bea
\Tr(XY^2) = \frac{1}{2} \Big(2 m_2 m_4 + m_1 m_5 - m_1 m_2^2 \Big).
\label{frstempl}
\eea
Swapping $X$ and $Y$ gives  
\bea
\Tr(YX^2) = \frac{1}{2} \Big(2 m_1 m_4 + m_2 m_3 - m_1^2 m_2 \Big).
\label{scndempl}
\eea
As $m+n$ increases, the number of distinct gauge-invariant operators that can be constructed increases. It is necessary to establish that a sufficient number of independent trace relations exist to express all such operators in terms of the primary invariants. A useful illustrative example is the case $m=2=n$. Two independent operators, $\Tr(X^2Y^2)$ and $\Tr(XYXY)$, can be constructed. The trace relation $T_2(Y^2, X, X) = 0$ gives  
\bea
\Tr(X^2Y^2) = \frac{1}{2} \Big(m_3 m_5 + 2 m_1 m_2 m_4 + m_1^2 m_5 - m_1^2 m_2^2 - m_1^2 m_5 \Big).
\label{thrdexmpl}
\eea
Similarly, $T_2(XY, X, Y) = 0$ after using (\ref{frstempl}), (\ref{scndempl}), and (\ref{thrdexmpl}), implies
\bea
\Tr(XYXY) = \frac{1}{2} \Big(m_2^2 m_3 + 2 m_4^2 - m_3 m_5 + m_1^2 m_5 - m_1^2 m_2^2 \Big).
\eea
Crucially, the growth in the number of independent operators is matched by the emergence of additional trace relations. To prove that this pattern holds in general, we now prove that all gauge invariant operators can be expressed in terms of the primary invariants (\ref{mesonloopsN2M2}) using trace relations. The proof proceeds by induction.

Assume all single-trace loops containing at most $k$ matrices are determined by the trace relations. We have already established this result for $k \leq 4$. Now, consider the loop $\Tr(X^{n_1}Y^{m_1})$ with $n_1 + m_1 = k + 1$ for $k \geq 4$. At least one of $n_1$ or $m_1$ must be greater than 1. Without loss of generality, assume $n_1 > 1$. The trace relation for $A = X$, $B = X^{n_1-1}$, and $C = Y^{m_1}$ is  
\bea
&&2\Tr(X^{n_1}Y^{m_1}) - \Tr(X) \Tr(X^{n_1-1}Y^{m_1}) - \Tr(X^{n_1}) \Tr(Y^{m_1}) \cr\cr
&&- \Tr(XY^{m_1}) \Tr(X^{n_1-1}) + \Tr(X) \Tr(X^{n_1-1}) \Tr(Y^{m_1}) = 0.
\eea
By the induction hypothesis, every term in this equation except the first contains at most $k$ matrices in the trace and is thus expressible in terms of the primary invariants (\ref{mesonloopsN2M2}). This establishes that $\Tr(X^{n_1}Y^{m_1})$ can also be expressed in terms of these variables. The same argument applies, with trivial changes, in the case where $m_1 > 1$.  

Next, consider invariants $\Tr(X^{n_1}Y^{m_1} \cdots X^{n_q}Y^{m_q})$, with  
\bea
n_1 + m_1 + \cdots + n_q + m_q = k+1.
\eea
Refer to invariants, with $q$ alternating blocks of $X^\#Y^\#$, as type-$q$ invariants. The trace relation obtained from $A = X^{n_1}$, $B = Y^{m_1}$, and $C = X^{n_2}Y^{m_2} \cdots X^{n_q}Y^{m_q}$, is  
\bea
&&\Tr(X^{n_1}Y^{m_1} \cdots X^{n_q}Y^{m_q+m_1}) + \Tr(X^{n_1+n_2}Y^{m_2} \cdots X^{n_q}Y^{m_q+m_1}) \cr\cr
&&- \Tr(X^{n_1}) \Tr(X^{n_2}Y^{m_2} \cdots X^{n_q}Y^{m_q+m_1}) - \Tr(Y^{m_1}) \Tr(X^{n_1+n_2}Y^{m_2} \cdots X^{n_q}Y^{m_q}) \cr\cr
&&- \Tr(X^{n_1}Y^{m_1}) \Tr(X^{n_2}Y^{m_2} \cdots X^{n_q}Y^{m_q}) 
+ \Tr(X^{n_1}) \Tr(Y^{m_1}) \Tr(X^{n_2}Y^{m_2} \cdots X^{n_q}Y^{m_q}) = 0.
\nonumber
\eea
Terms on the second and third lines contain at most $k$ matrices in their traces and are, by the induction hypothesis, expressible in terms of the primary invariants (\ref{mesonloopsN2M2}). The first term on the first line is the desired type-$q$ invariant, while the second term on the first line is a type-$(q-1)$ invariant.  Since we have already established that the type-$1$ invariant $\Tr(X^{n_1}Y^{m_1})$ can be expressed in terms of the primary invariants, applying the above trace relation proves that the type-$2$ invariant $\Tr(X^{n_1}Y^{m_1}X^{n_2}Y^{m_2})$ can also be expressed in terms of these variables. This reasoning extends recursively, proving that all type-$q$ invariants can be determined in terms of the primary invariants.

Thus, assuming all single-trace invariants with at most $k$ matrices can be expressed in terms of the primary invariants (\ref{mesonloopsN2M2}) using trace relations, we have established that all single-trace invariants with $k+1$ matrices can also be expressed in these terms. This completes the proof that, for $N=2$, the complete space of gauge invariants can be written in terms of the primary invariants.

\subsubsection{Three matrix model}

Consider a free matrix model with three species of matrices: $X$, $Y$, and $Z$. The partition function is expressed in terms of the variables  
\bea
x &=& e^{-\beta E_1 - \mu_x}, \qquad  
y \,\,=\,\, e^{-\beta E_2 - \mu_y}, \qquad  
z \,\,=\,\, e^{-\beta E_3 - \mu_z},
\eea
where the chemical potentials $\mu_x$, $\mu_y$, and $\mu_z$ keep track of the three matrices. The Molien-Weyl formula gives the graded partition function
\bea
Z(x,y,z) = \frac{1 + xyz}{(1-x)(1-y)(1-z)(1-x^2)(1-y^2)(1-z^2)(1-xy)(1-xz)(1-yz)}.\cr\label{exactN2M3}
\eea
A new feature compared to the two-matrix model is the monomial $xyz$ appearing with a positive sign in the numerator. It corresponds to a secondary invariant. The coefficient of $x^n y^m z^p$ in the power series expansion of this partition function, counts the number of gauge invariants that can be constructed using $n$ $X$s, $m$ $Y$s, and $p$ $Z$s.  
Using \texttt{Mathematica}, we have verified the series expansion of (\ref{exactN2M3}) perfectly matches (\ref{LRPFMM}) for all invariants constructed using up to 30 matrices.

This partition function suggests that the complete set of gauge-invariant operators is generated by the following nine primary invariants and a single secondary invariant
\bea
m_1 &=& \Tr(X), \qquad m_2 = \Tr(Y), \qquad m_3 = \Tr(Z), \cr\cr
m_4 &=& \Tr(X^2), \qquad m_5 = \Tr(Y^2), \qquad m_6 = \Tr(Z^2), \cr\cr
m_7 &=& \Tr(XY), \qquad m_8 = \Tr(YZ), \qquad m_9 = \Tr(ZX), \qquad
s = \Tr(XYZ). \label{loopsN2M3}
\eea
The primary invariants are again given by the complete set of single trace operators with $\le N=2$ matrices in the trace. The partition function (\ref{exactN2M3}) and the generators (\ref{loopsN2M3}) were independently derived in \cite{3mats} using mathematical methods based on graded rings. 

The understand why the secondary invariant $s$ is needed, consider the single-trace loops containing three matrices. If all three matrices belong to the same species, there is a single gauge-invariant operator, determined by a single trace relation. For instance, $\Tr(X^3)$ is fixed by the trace relation obtained from $A = B = C = X$. Similarly, if the three matrices belong to only two species, there is again a single gauge-invariant operator determined by a single trace relation, e.g., the expression for $\Tr(X^2 Y)$ follows by choosing $A = B = X$ and $C = Y$. When all three matrix species appear, there are two independent single-trace operators, $\Tr(XYZ)$ and $\Tr(XZY)$, but still only one trace relation. We choose to retain $\Tr(XYZ)$ as an independent variable and express $\Tr(XZY)$ in terms of it using the trace relation. Alternatively, we could have kept $\Tr(XZY)$ and solved for $\Tr(XYZ)$. This demonstrates the necessity of introducing the secondary invariant.

Secondary invariants are quadratically reducible. Thus they obey a relation which expresses $s^2$ as a polynomial in the primary and secondary invariants, with the secondary invariant $s$ appearing at most linearly. The relevant constraint, which is easily verified numerically, is:
\bea
&&s^2 + s \left(m_{1} m_{2} m_{3} - m_{1} m_{8} - m_{2} m_{9} - m_{3} m_{7} \right) 
+ \frac{1}{4} m_{1}^2 m_{2}^2 m_{3}^2 
- \frac{1}{2} m_{1}^2 m_{2} m_{3} m_{8} 
- \frac{1}{4} m_{1}^2 m_{5} m_{6} \cr\cr
&&+ \frac{m_{1}^2 m_{8}^2}{2} 
- \frac{1}{2} m_{1} m_{2}^2 m_{3} m_{9} 
- \frac{1}{2} m_{1} m_{2} m_{3}^2 m_{7} 
+ \frac{1}{2} m_{1} m_{2} m_{6} m_{7} 
+ \frac{1}{2} m_{1} m_{3} m_{5} m_{9} 
- \frac{1}{4} m_{2}^2 m_{4} m_{6} \cr\cr
&&+ \frac{m_{2}^2 m_{9}^2}{2} 
+ \frac{1}{2} m_{2} m_{3} m_{4} m_{8} 
- \frac{1}{4} m_{3}^2 m_{4} m_{5} 
+ \frac{m_{3}^2 m_{7}^2}{2} 
+ \frac{m_{4} m_{5} m_{6}}{2} 
- \frac{m_{4} m_{8}^2}{2} 
- \frac{m_{5} m_{9}^2}{2} 
- \frac{m_{6} m_{7}^2}{2} \cr\cr
&&+ m_{7} m_{8} m_{9} = 0.
\eea

To complete our discussion, we now prove that all gauge-invariant operators can be expressed in terms of the invariants (\ref{loopsN2M3}) using trace relations. Our proof proceeds by induction.

Assume that all single-trace operators containing at most $k$ matrices can be written in terms of the invariants (\ref{loopsN2M3}). Consider a typical operator of the form
\bea
\phi_t \equiv \Tr(X^{n_1}Y^{m_1}Z^{p_1}X^{n_2}Y^{m_2}Z^{p_2} \cdots X^{n_q}Y^{m_q}Z^{p_q}).
\eea
with
\bea
n_1+m_1+p_1+\cdots+n_q+m_q+p_q=k+1
\eea
Suppose at least one of the integers $\{n_1, m_1, p_1, \dots, n_q, m_q, p_q\}$ is greater than 1. Without loss of generality, assume $p_1 > 1$. The trace relation obtained by setting $A = Z$, $B = Z^{p_1 - 1}$, and 
\bea
C = X^{n_2}Y^{m_2}Z^{p_2} \cdots X^{n_q}Y^{m_q}Z^{p_q}X^{n_1}Y^{m_1}
\nonumber
\eea
yields
\bea
&&2\phi_t - \Tr(Z) \Tr(Z^{p_1-1} C) - \Tr(Z^{p_1-1}) \Tr(Z C) \cr\cr
&&-\Tr(Z^{p_1}) \Tr(C) + \Tr(Z^{p_1-1}) \Tr(Z) \Tr(C) \,\,=\,\, 0.
\eea
By the induction hypothesis, all terms in this equation except for $\phi_t$ contain at most $k$ matrices in the trace and can thus be expressed in terms of the variables (\ref{loopsN2M3}). This establishes that $\phi_t$ is also determined by these variables.

To complete the proof, we must consider operators in which no matrix is raised to a power greater than 1. One such family is given by
\bea
\phi_{a1} = \Tr\Big((XYZ)^k\Big).
\eea
For $k=1$, this is simply the secondary invariant $s$, so there is nothing to prove. For $k > 1$, choosing $A = XY$, $B = Z$, and $C = (XYZ)^{k-1}$ leads to the trace relation
\bea
&&\phi_{a1} + \Tr\Big(XY(XYZ)^{k-1}Z\Big) - \Tr(XY) \Tr\Big(Z(XYZ)^{k-1}\Big) \cr\cr
&&- \Tr(Z) \Tr\Big(XY(XYZ)^{k-1}\Big) - \Tr(XYZ) \Tr\Big((XYZ)^{k-1}\Big) \cr\cr
&&+ \Tr(Z) \Tr(XY) \Tr\Big((XYZ)^{k-1}\Big) \,\,=\,\, 0.
\eea
The second term on the left-hand side is a typical operator, while the remaining terms contain at most $k$ matrices in the trace and are therefore, by the induction hypothesis, expressible in terms of (\ref{loopsN2M3}). Consequently, the atypical operator $\phi_{a1}$ can also be expressed in terms of these variables.

A completely analogous argument applies to the following atypical operators:
\bea
\phi_{a2} = \Tr\Big((XY)^k\Big), \qquad \phi_{a3} = \Tr\Big((XZ)^k\Big), \qquad \phi_{a4} = \Tr\Big((YZ)^k\Big).
\eea
Finally, the same reasoning extends to the most general atypical operator:
\bea
\phi_a = \Tr\big( (XYZ)^{k_1} (XY)^{l_1} (XZ)^{m_1} (YZ)^{n_1} (XYZ)^{k_2} \cdots \big).
\eea
By partitioning the sequence of matrices being traced into three factors, $A$, $B$, and $C$, where $A$ and $B$ are chosen such that $\Tr(ABC)$ corresponds to the atypical operator and $\Tr(ACB)$ is a typical operator, we can apply the same trace relations to show that $\phi_a$ is also determined by (\ref{loopsN2M3}).

This completes the proof that in the three-matrix model at $N=2$, the complete space of gauge-invariant operators can be written in terms of the variables (\ref{loopsN2M3}).

\subsubsection{Four matrix model}\label{fourmatrixdiscusion}

To study the four-matrix model constructed using matrices $W$, $X$, $Y$ and $Z$, introduce the variables
\bea
w &=& e^{-\beta E_1 - \mu_w}, \qquad
x \,\,=\,\, e^{-\beta E_2 - \mu_x}, \qquad
y \,\,=\,\, e^{-\beta E_3 - \mu_y}, \qquad  
z \,\,=\,\, e^{-\beta E_4 - \mu_z},
\eea
The graded partition function is
\bea
Z(w, x, y, z) &=& \frac{1}{(1 - w)(1 - x)(1 - y)(1 - z)(1 - w^2)(1 - x^2)(1 - y^2)(1 - z^2)}\cr\cr
&\times& \frac{P(w, x, y, z)}{(1 - xy)(1 - xw)(1 - xz)(1 - wy)(1 - wz)(1 - yz)}, \label{N2M4}
\eea
where
\bea
P(w, x, y, z) &=& 1 + wxy + wxz + wyz + xyz - w^2 xyz - wx^2 yz \cr\cr
&& - wxy^2 z - wxyz^2 - w^2 x^2 y^2 z^2.
\eea
This is the most general partition function obtained so far. Each factor in the denominator corresponds to a primary invariant, giving a total of 14 such invariants. The monomials in the numerator with positive coefficients represent secondary invariants, of which there are four. Terms with negative coefficients signal the presence of relations among the invariants, indicating that the full ring of gauge-invariant operators is not freely generated. These negative terms correct for the overcounting that results from treating all generators as independent. To validate the result (\ref{N2M4}), we have confirmed that its series expansion agrees with (\ref{LRPF}) for all gauge-invariant operators built from up to 30 matrices.

The Hilbert series is a rational function for a broad class of rings, and as in our example, negative terms in the numerator reflect the presence of constraints among the generators. In more general settings, an additional feature may arise: the existence of constraints among the constraints themselves. These higher-order relations are known as \emph{syzygies}. Their presence implies that imposing only the primary constraints leads to an undercounting of gauge-invariant operators \cite{MW1}. To compensate, additional higher-degree terms with positive coefficients must be included in the numerator of the partition function. These syzygies can themselves satisfy further relations, called \emph{higher syzygies}, which introduce new terms in the numerator, this time with negative coefficients. This hierarchical structure of relations can continue through multiple levels, with each successive syzygy layer modifying the numerator alternately with positive and negative contributions. The fact that this sequence of corrections eventually terminates is guaranteed by Hilbert's Syzygy Theorem \cite{HST}.

From the graded partition function we read of the following primary and secondary invariants
\bea
m_1 &=& \Tr(W), \qquad m_2 = \Tr(X), \qquad m_3 = \Tr(Y), \qquad m_4 = \Tr(Z),\cr\cr
m_5 &=& \Tr(W^2), \qquad m_6 = \Tr(X^2), \qquad m_7 = \Tr(Y^2), \qquad m_8 = \Tr(Z^2),\cr\cr
m_9 &=& \Tr(WX), \qquad m_{10} = \Tr(WY), \qquad m_{11} = \Tr(WZ), \qquad m_{12} = \Tr(XY),\cr\cr
m_{13} &=& \Tr(XZ), \qquad m_{14} = \Tr(YZ), \qquad
s_1 = \Tr(WXY), \qquad s_2 = \Tr(WXZ),\cr\cr
s_3 &=& \Tr(WYX), \qquad s_4 = \Tr(XYZ). \label{invtvarsN2M4}
\eea
We can confirm the validity of the proposed invariants through an algebraic analysis based on trace relations. To understand the necessity of introducing the four secondary invariants, consider single-trace operators built from three matrices. When all three matrices are of the same species, or drawn from only two species, there exists a unique gauge-invariant operator determined by a single trace relation. However, when the three matrices are of different species, two independent gauge-invariant operators can be constructed. For example, both $\Tr(WXY)$ and $\Tr(WYX)$ are valid gauge-invariant operators. The trace relation $T_2(W, X, Y) = 0$ ensures that one of these can be written in terms of the other, leaving one undetermined. This undetermined operator must be included as a secondary invariant. As discussed earlier, there is some flexibility in choosing which of the two to treat as a secondary invariant and which to eliminate using the trace relations. To establish that these additional invariants are indeed secondary, we must identify constraints that express the product of two such secondary invariants in terms of polynomials involving primary and secondary invariants, with at most linear dependence on the secondary invariants. By numerically searching for combinations of invariants that sum to zero, we find ten such relations, which are summarized in Appendix \ref{solitonsN2M4}.

The negative terms in the numerator of (\ref{N2M4}) reflect constraints among the invariants (\ref{invtvarsN2M4}). The grading of each negative term specifies which matrices participate in a given constraint. For example, the term $-w^2xyz$ indicates a constraint involving two $W$s, an $X$, a $Y$ and a $Z$. A numerical search confirms the existence of such a nontrivial relation:
\bea
&&-\frac{1}{2}m_1^2m_{13}m_3+\frac{1}{2}m_1^2s_4+\frac{1}{2}m_1m_{10}m_{13} 
-\frac{1}{2}m_1m_{11}m_{12}+\frac{1}{2}m_1m_{11}m_2m_3-\frac{1}{2}m_1m_{14} m_9\cr\cr
&& -\frac{1}{2}m_1 m_2 s_3+\frac{1}{2} m_1m_3m_4m_9+\frac{1}{2}m_1m_3s_2 
-\frac{1}{2}m_1m_4s_1-m_{10}s_2- m_{11} m_{3} m_{9} + m_{11} s_{1} \cr\cr
&&+\frac{1}{2}m_{12}m_4m_5+\frac{1}{2}m_{13}m_3m_5+\frac{1}{2}m_{14} m_2 m_5 
-\frac{1}{2} m_{2} m_{3} m_{4} m_{5} - m_{5} s_{4} + m_{9} s_{3} \,\,=\,\,0.\cr
&&
\eea
We have verified that there is a constraint for each negative term in the numerator of (\ref{N2M4}). The complete set of constraints are recorded in Appendix \ref{constraintsN2M4}. This confirms the interpretation of the partition function (\ref{N2M4}) and verifies our set of invariants (\ref{invtvarsN2M4}). 

Our partition function does not initially appear in the expected form (\ref{illpf}) associated with a Hironaka decomposition. To proceed, switch off the chemical potentials for each matrix species by setting $x = w = y = z$, which defines a simplified grading. With this grading, the partition function becomes
\begin{align}
Z(x) = \frac{1 + x^2 + 4x^3 + x^4 + x^6}{(1 - x)^4 (1 - x^2)^9},
\end{align}
which clearly exhibits the Hironaka form. This confirms that the Hironaka decomposition does not respect a grading by individual matrix species. Instead, it is compatible only with the total-degree grading.

Fortunately, the complete set of primary and secondary invariants for this ring has been constructed in \cite{Teranishi39,Domokos}, allowing us to verify this conclusion explicitly. The primary invariants are
\begin{align*}
m_1 &= \Tr(W), & m_2 &= \Tr(X), & m_3 &= \Tr(Y), & m_4 &= \Tr(Z), \\
m_5 &= \Tr(W^2), & m_6 &= \Tr(X^2), & m_7 &= \Tr(Y^2), & m_8 &= \Tr(Z^2), \\
m_9 &= \Tr(WX), & m_{10} &= \Tr(WY), & m_{11} &= \Tr(WZ) + \Tr(XY), \\
m_{12} &= \Tr(XZ), & m_{13} &= \Tr(YZ).
\end{align*}
Note that there are $(d - 1)N^2+1=13$ primary invariants, matching the Krull dimension of the ring. This confirms that they are algebraically independent, with no relations among them. By contrast, the fully graded partition function featured 14 generators, implying the existence of relations - as we confirmed earlier.

The secondary invariants are given by
\begin{align*}
s_1 &= \Tr(WZ), & s_2 &= \Tr(WXY), & s_3 &= \Tr(WXZ), & s_4 &= \Tr(WYX), \\
s_5 &= \Tr(XYZ), & s_6 &= \Tr(WZ)^2, & s_7 &= \Tr(WZ)^3.
\end{align*}
This result is particularly revealing: the invariant $m_{11}=\Tr(WZ)+\Tr(XY)$ is not homogeneous under the matrix-species grading. This is precisely why the Hironaka decomposition fails to align with that grading.

This observation offers an important lesson: to verify a Hironaka decomposition, one should always consider the simply graded partition function (graded by total degree), rather than attempting to preserve matrix-species grading. In fact, \cite{Djokovic} proves that no bigraded system of primary invariants exists for the rings $C_{2,5}$ and $C_{2,6}$, implying that their Hironaka decompositions also do not respect a matrix-species grading.

\subsection{$N=3$}

The analysis at $N=3$ is considerably more intricate than at $N=2$. Owing to the increased complexity, we limit our study to the two- and three-matrix models. For the two-matrix model, the bigraded partition function shows that the full set of gauge-invariant operators is freely generated, with both primary and secondary invariants contributing to the spectrum.

In contrast, the bigraded partition function for the three-matrix model does not exhibit the Hironaka form. This motivates a computation of the simply graded partition function, which does take the expected Hironaka form. This result proves that no trigraded system of primary invariants exists for the ring $C_{3,3}$.

\subsubsection{Two matrix model}\label{2matn3discuss}

The graded partition function is given by  
\bea
Z(x) &=& \frac{1 + x^3y^3}{(1-x)(1-y)(1-x^2)(1-xy)(1-y^2)(1-x^3)(1-x^2y)(1-xy^2)(1-y^3)(1-x^2y^2)}.\cr\cr
&&\label{N3M2}
\eea
We have verified that the series expansion of (\ref{N3M2}) is in complete agreement with (\ref{LRPF}) for all invariants constructed using up to 30 matrices.

The result (\ref{N3M2}) implies that the ring is freely generated and that the complete set of primary invariants are those given in (\ref{mesonloopsN2M2}) along with  
\bea
m_6 &=& \Tr(X^3), \qquad m_7 = \Tr(X^2Y), \qquad m_8 = \Tr(XY^2), \qquad m_9 = \Tr(Y^3),\cr\cr
m_{10} &=& \Tr(X^2Y^2),
\label{mesonsN3M2}
\eea
and there is a single secondary invariant,  
\bea
s &=& \Tr(XYX^2Y^2).
\label{solitonN3M2}
\eea
Notice that primary invariants with more than $N$ matrices in the trace have appeared for the first time. The partition function (\ref{N3M2}) and this set of generators agree with the results of \cite{3mats}.

We can again verify these results algebraically using the trace relations. For $N=3$, the trace relations take the form $T_3(A,B,C,D)=0$ for any words $A$, $B$, $C$, and $D$, where the explicit expression for $T_3(A,B,C,D)$ is given in Appendix \ref{TrRelN3}.  

All loops involving at most three matrices are included among the primary invariants. To justify the inclusion of the invariant with four matrices in the trace, note that there are six single trace operators that can be constructed using four matrices. There are only five independent trace relations among them. As explained in Appendix \ref{TrRelN3}, the operator $\Tr(X^2Y^2)$ is not determined by the trace relations\footnote{There is again some freedom: we could choose $\Tr(XYXY)$ as a primary invariant, in which case the trace relations determine $\Tr(X^2Y^2)$.}. This justifies its inclusion as an additional invariant. It is natural to include it as a primary invariant since there are $(d-1)N^2+1=10$ primary invariants. Further, the above five trace relations exhaust all relations that can be formed with four matrices in the trace, so that the primary invariants are algebraically independent.

Among the six independent single-trace invariants that can be formed with five matrices, all can be expressed in terms of the primary invariants (\ref{mesonsN3M2}) using the trace relations. The explicit expressions are provided in Appendix \ref{TrRelN3}. Similarly, among the fourteen independent single-trace operators with six matrices, thirteen can be determined in terms of the primary invariants, leaving one undetermined operator: $s=\Tr(XYX^2Y^2)$. This operator must be included as an additional invariant. It is a secondary invariant due to the existence of the quadratic relation:  
\bea
s_1^2 + p_1(m_i)s_1 + p_2(m_i) = 0,
\eea
where $p_1(m_i)$ and $p_2(m_i)$ are polynomials in the primary invariants and are recorded in Appendix \ref{solrel}.

As the number of matrices increases, the number of trace relations grows. This growth parallels the increase in gauge-invariant operators with arbitrary trace structures, which outpaces the growth of single-trace operators. Consequently, while some number of secondary invariants may initially be necessary, when there are a sufficiently large number of matrices in the trace, the trace relations are always abundant enough to completely determine the complete set of gauge invariant operators.

\subsubsection{Three matrix model}

The graded partition function is a lengthy expression, which we do not explicitly quote here. The numerator is a polynomial with both positive and negative terms, and there are 22 factors in the denominator. This is not of the Hironaka form (\ref{illpf}). Setting the chemical potentials for the three species of matrices equal, we obtain the partition function, graded only by degree, given by  
\bea
Z(x) &=& \frac{P_{N=3,M=3}(x)}{(1-x)^3(1-x^2)^6(1-x^3)^8(1-x^4)^2},\label{N3M3}
\eea
where the numerator takes the form  
\bea
P_{N=3,M=3}(x) &=& 1+3 x^3+7 x^4+9 x^5+16 x^6+18 x^7+25 x^8+30 x^9+34 x^10+30 x^{11}\cr\cr
&&+25 x^{12}+18 x^{13}+16 x^{14}+9 x^{15}+7 x^{16}+3 x^{17}+x^{20}\,.
\eea
The series expansion of (\ref{N3M3}) is in complete agreement with (\ref{LRPF}) for all invariants constructed using up to 30 matrices. This partition function confirms the presence of $19=(d-1)N^2+1$ primary invariants, matching the Krull dimension as it must. There are 251 secondary invariants. We have not attempted to explicitly construct these invariants.

\subsection{$N=4$ to $N=7$}

For $N>3$, the partition functions are lengthy expressions. In this section, we simply evaluate the partition function, graded by degree. One could take the extra step of a detailed algebraic study with input from the trace relations. This becomes prohibitively tedious beyond a certain point. For example, at $N=4$, there are 63 secondary invariants. In principle, the trace relations could be used to confirm that the trace relations do not determine these gauge-invariant operators. Furthermore, the trace relations could also be used to verify the 2016 relations which would verify that these invariants are quadratically reducible. However, such a verification is impractical due to its sheer complexity.

Our objective in this section is to evaluate the two-matrix partition function to determine the number of primary and secondary invariants as a function of $N$.

\subsubsection{$N=4$}

From the Molien-Weyl formula  (\ref{MWPF}) we find the partition function
\bea
Z(x)&=&{P_{N=4,M=2}(x)\over (1-x)^2(1-x^2)^3(1-x^3)^4(1-x^4)^6(1-x^6)^2}\label{N4M2}
\eea
where
\bea
P_{N=4,M=2}(x)&=&1 + 2 x^5 + 2 x^6 + 2 x^7 + 4 x^8 + 4 x^9 + 4 x^{10} 
+ 4 x^{11} + 
 2 x^{12} + 4 x^{13} + 4 x^{14}\cr\cr
&& + 4 x^{15} + 4 x^{16} + 2 x^{17}+2 x^{18}+2 x^{19}+x^{24}
\eea
We have verified that the series expansion of (\ref{N3M2}) is in complete agreement with (\ref{LRPF}) for all invariants constructed using up to 30 matrices. This partition function was first obtained in \cite{Teranishi}. The partition function suggests that the space of gauge invariants admits a Hironaka decomposition with a total of 17 primary and 63 secondary invariants. 

\subsubsection{$N=5$}

The Molien-Weyl formula (\ref{MWPF}) gives the blind partition function
\bea
Z(x)&=&\frac{P_{N=5,M=2}(x)}{(1-x)^2 \left(1-x^2\right)^3 \left(1-x^3\right)^4 \left(1-x^4\right)^6 \left(1-x^5\right)^6 \left(1-x^6\right)^5},\label{N5M2}
\eea
where
\bea
P_{N=5,M=2}(x)&=&1 + 2 x^5 + 2 x^6 + 8 x^7 + 13 x^8 + 16 x^9 + 25 x^{10} + 28 x^{11} + 
 46 x^{12} + 58 x^{13} + 85 x^{14}\cr\cr
&&+132 x^{15}+172 x^{16}+232 x^{17}+282 x^{18}+346x^{19}+404 x^{20}+444 x^{21}+518 x^{22}\cr\cr
&&+570 x^{23}+633x^{24}+684 x^{25}+711 x^{26}+744 x^{27}+711 x^{28}+684 x^{29}+633 x^{30}\cr\cr
&&+570x^{31}+518 x^{32}+444 x^{33}+ 404 x^{34}+ 346 x^{35} + 
 282 x^{36} + 232 x^{37} + 172 x^{38} \cr\cr
&& + 132 x^{39} + 85 x^{40} + 58 x^{41} + 
 46 x^{42} + 28 x^{43}+ 25 x^{44}+ 16 x^{45} + 13 x^{46}\cr\cr
&&+ 8 x^{47} + 2 x^{48} + 2 x^{49} + x^{54}.
\eea
The series expansion of (\ref{N5M2}) is in complete agreement with (\ref{LRPF}) for all invariants constructed using up to 30 matrices. This partition function was first obtained in \cite{Djokovic}. When comparing formulae for parition functions, common factors in the numerator and denominator may need to be cancelled before two formulas manifestly agree. 
For example, the denominator of the partition function given in \cite{Djokovic} is:
\bea
(1-x^2)^6(1- x^3)^8(1-x^4)^6 (1-x^5)^6.
\eea
There are 26 factors, which agrees with (\ref{N5M2}). However, interpreting the above expression literally would suggest there are 6 primary invariants of length 2, 8 of length 3, 6 of length 4, and 6 of length 5. This is inconsistent, as there are only 2 independent single trace operators of length 2 and 4 independent single trace operators of length 8. Cancelling common factors between numerator and denominator in our result reduces it to that of \cite{Djokovic}. The partition function (\ref{N5M2}) implies that the space of gauge invariants admits a Hironaka decomposition with a total of 26 primary invariants and 15,423 secondary invariants. 

A new effect, that first appears for $N=5$ and continues for all higher $N$, is that not all single trace operators with $\le N=5$ matrices in the trace are included as primary invariants. The second term in the numerator makes it clear that two loops of length 5 are included as secondary invariants. We comment further on this behaviour in Section \ref{NumbMes}.

\subsubsection{$N=6$}

For $N=6$ the Molien-Weyl formula  (\ref{MWPF}) yields the partition function
\bea
Z={P_{N=6,M=2}(x)\over (1-x)^2(1-x^2)^3(1-x^3)^4(1-x^4)^6(1-x^5)^7(1-x^6)^9(1-x^8)^6}\label{N6M2}
\eea
where
\bea
&&P_{N=6,M=2}(x)=1+x^5+5 x^6+12 x^7+16 x^8+32 x^9+53 x^{10}+77 x^{11}+143 x^{12}+228 x^{13}\cr\cr
&&\quad +392 x^{14}+645 x^{15}+1073 x^{16}+1707 x^{17}+2707 x^{18}+4236 x^{19}+6431 x^{20}+9741 x^{21}\cr\cr
&&\quad +14487 x^{22}+21227 x^{23}+30799 x^{24}+44067 x^{25}+62207 x^{26}+86643 x^{27}+119251 x^{28}\cr\cr
&&\quad +161759 x^{29}+216606 x^{30}+286615 x^{31}+373964 x^{32}+482231 x^{33}+614103 x^{34}\cr\cr
&&\quad +772510 x^{35}+959988 x^{36}+1178996 x^{37}+1430764 x^{38}+1715471 x^{39}+2033523 x^{40}\cr\cr
&&\quad +2381900 x^{41}+2757697 x^{42}+3156544 x^{43}+3571370 x^{44}+3994807 x^{45}+4418149 x^{46}\cr\cr
&&\quad +4831561 x^{47}+5223944 x^{48}+5585650 x^{49}+5906278 x^{50}+6174865 x^{51}+6385150 x^{52}\cr\cr
&&\quad +6529068 x^{53}+6601986 x^{54}+6601986 x^{55}+6529068 x^{56}+6385150 x^{57}+6174865 x^{58}\cr\cr
&&\quad +5906278 x^{59}+5585650 x^{60}+5223944 x^{61}+4831561 x^{62}+4418149 x^{63}+3994807 x^{64}\cr\cr
&&\quad +3571370 x^{65}+3156544 x^{66}+2757697 x^{67}+2381900 x^{68}+2033523 x^{69}+1715471 x^{70}\cr\cr
&&\quad +1430764 x^{71}+1178996 x^{72}+959988 x^{73}+772510 x^{74}+614103 x^{75}+482231 x^{76}\cr\cr
&&\quad +373964 x^{77}+286615 x^{78}+216606 x^{79}+161759 x^{80}+119251 x^{81}+86643 x^{82}+62207 x^{83}\cr\cr
&&\quad +44067 x^{84}+30799 x^{85}+21227 x^{86}+14487 x^{87}+9741 x^{88}+6431 x^{89}+4236 x^{90}+2707 x^{91}\cr\cr
&&\quad +1707 x^{92}+1073 x^{93}+645 x^{94}+392 x^{95}+228 x^{96}+143 x^{97}+77 x^{98}+53 x^{99}+32 x^{100}\cr\cr
&&\quad +16 x^{101}+12 x^{102}+5 x^{103}+x^{104}+x^{109}
\eea
We have checked that the series expansion of (\ref{N6M2}) is in complete agreement with (\ref{LRPF}) for all invariants constructed using up to 30 matrices. This partition function was first obtained in \cite{Djokovic}. The partition function proves that the space of gauge invariants is generated with a total of 37 primary invariants and 312,606,719 secondary invariants. 

\subsubsection{$N=7$}

Using the Molien-Weyl formula  (\ref{MWPF}) we find the partition function
\bea
Z={P_{N=7,M=2}(x)\over (1-x)^2(1-x^2)^3(1-x^3)^4(1-x^4)^6(1-x^5)^8(1-x^6)^{11}(1-x^7)^8(1-x^8)^5(1-x^{10})^2(1-x^{12})}\cr\label{N7M2}
\eea
where $P_{N=7,M=2}(x)$ is a polynomial of degree 180, given in Appendix \ref{N7M2PF}. Again, we have verified that the series expansion of (\ref{N7M2}) is in complete agreement with (\ref{LRPF}) for all invariants constructed using up to 30 matrices. This partition function was first obtained in \cite{Kristensson:2020nly}. The partition function proves that the space of gauge invariants is generated using 50 primary invariants and a staggering 21,739,438,196,735 secondary invariants. 

$P_{N=7,M=2}(x)$ starts as $1+3 x^6+\cdots$ so that there are no longer any secondary invariants of length 5. This is a general trend: as we increase $N$ low lying states that were represented as secondary invariants now appear as primary invariants.

\subsection{General $N$}

In this section, we argue that the results obtained above already point to several general patterns. Specifically, the partition functions we have evaluated exhibit temperature inversion symmetry. They all involve a total of $(d-1)N^2+1$ primary invariants and a number of secondary invariants that appears to grow as $\sim e^{c N^2}+\cdots$ where $c$ is a constant and $\cdots$ are terms that are sub leading at large $N$.

\subsubsection{Temperature inversion symmetry}\label{tempinversion}

The numerators of our partition functions exhibit a palindromic structure. A palindromic polynomial is one whose coefficients are symmetric, meaning they remain unchanged when read forward and backward. More concretely, a polynomial
\bea
P(x)=a_0+a_1 x+a_2 x^2 +\cdots +a_n x^n
\eea
is palindromic if
\bea
a_k=a_{n-k}, \qquad \text{for all } k=0,1,\dots,n.
\eea
For example, the polynomial $x^4 +3x^3+5x^2 +3x+1$ is palindromic. The palindromic nature of our partition functions was originally established in \cite{3mats} using the properties of the Molien-Weyl partition function. Specifically, Proposition 2.3 in \cite{3mats} proves this result for the graded partition function relevant to any number of matrix species and any $N$. See also \cite{Kristensson:2020nly}. In the case of two $N\times N$ matrices, the relevant relation is
\bea
Z_N(1/x_1,1/x_2)&=&(-1)^{N-1} (x_1x_2)^{N^2} Z_N(x_1,x_2).
\eea
The key idea behind this proof is intuitive. The Molien-Weyl partition function involves an integral over the unit circle in the complex plane. The transformation $x_1 \to 1/x_1$ and $x_2 \to 1/x_2$ exchanges the poles inside and outside the unit circle. By Cauchy's integration theorem, the sum of all residues must vanish, implying that the sum of residues from poles outside the unit circle is, up to a sign, equal to that from poles inside.

For the degree graded partition function, obtained by setting $x_1 = x_2 = x$, this reduces to
\bea
Z_N(1/x)&=&(-1)^{N-1} x^{2N^2} Z_N(x).
\eea
We can transform this relation into an exact symmetry by defining
\bea
\tilde{Z}_N(x) = x^{N^2} Z_N(x).
\eea
The new function satisfies
\bea
\tilde{Z}_N(1/x) &=& %x^{-N^2} Z_N(1/x) \cr\cr
%&=& (-1)^{N-1} x^{-N^2} x^{2N^2} Z_N(x) \cr\cr
%&=& (-1)^{N-1} x^{N^2} Z_N(x) \,=\, 
(-1)^{N-1} \tilde{Z}_N(x),
\eea
demonstrating that, up to a phase, $\tilde{Z}_N(x)$ is invariant under the transformation $x\to 1/x$.

This property has a natural physical interpretation. In the free theory, each matrix contributes $N^2$ oscillators, each with a ground state energy of ${1\over 2}$\footnote{In our partition function calculations, we assume an energy gap of $E=1$ between levels, leading to a ground state energy of ${1\over 2}$. See Section \ref{2M2N}.}. Consequently, the total ground state energy contribution from two matrices is given by
\bea
e^{-\beta 2N^2 {1\over 2}} = e^{-\beta N^2} = x^{N^2},
\eea
which precisely accounts for the transformation factor that converts $Z_N(x)$ into $\tilde{Z}_N(x)$. Thus, the palindromic nature of the partition function emerges as a direct consequence of the $x\to 1/x$ symmetry of the free partition function.
Since $x = e^{-\beta}$, the transformation $x\to 1/x$ corresponds to $\beta\to -\beta$, which is a temperature reflection, $T \to -T$. This symmetry is a known feature of finite-temperature path integrals in quantum field theory \cite{McGady:2017rzv}.

\subsubsection{Number of primary invariants}\label{NumbMes}

The partition functions discussed above are all of the Hironaka form which demonstrates that the complete set of gauge invariants is freely generated using a set of primary and secondary invariants. Denoting the number of primary invariants by $M$, the multi matrix models discussed above obey the relation
\bea
M=(d-1)N^2+1,
\eea
for all $N$.

All single-trace operators involving at most $N$ matrices must be included among the generating invariants. This is because these operators can't be written in terms of others: trace relations only begin to appear when the number of matrices in the trace reaches $N+1$. As a result, all single traces with $\leq N$ matrices are algebraically independent.

The \emph{length} $L$ of a single-trace operator refers to the number of matrices appearing in the trace. To estimate how the number of such operators grows with $L$, we follow the approach of \cite{Berenstein:2018hpl}. In a matrix model with $d$ matrix species, a single-trace operator of length $L$ is constructed by tracing a product of $L$ matrices, each chosen from the $d$ available species. A naive estimate of the number of such products is $d^L$.

However, this overcounts due to the cyclic symmetry of the trace. For example, the words $XYY$, $YXY$ and $YYX$ are distinct as sequences but yield the same trace: $\Tr(XYY) = \Tr(YXY) = \Tr(YYX)$. A more refined approximation for the number of distinct single-trace operators of length $L$, denoted $N_{\text{op}}(L)$, is
\begin{align}
N_{\text{op}}(L) \approx \frac{d^L}{L}.
\end{align}
This estimate, while useful, undercounts the true number of operators because it neglects additional symmetries within certain traces. For instance, the trace $\Tr(X^2 Y X^2 Y)$ corresponds to only three distinct words - $XXYXXY$, $XYXXYX$, and $YXXYXX$ - due to its internal symmetries. Thus, although $N_{\text{op}}(L)$ captures the leading behavior, it systematically underestimates the actual count of single-trace operators.

\begin{figure}[h]
\begin{center}
\includegraphics[width=0.9\columnwidth]{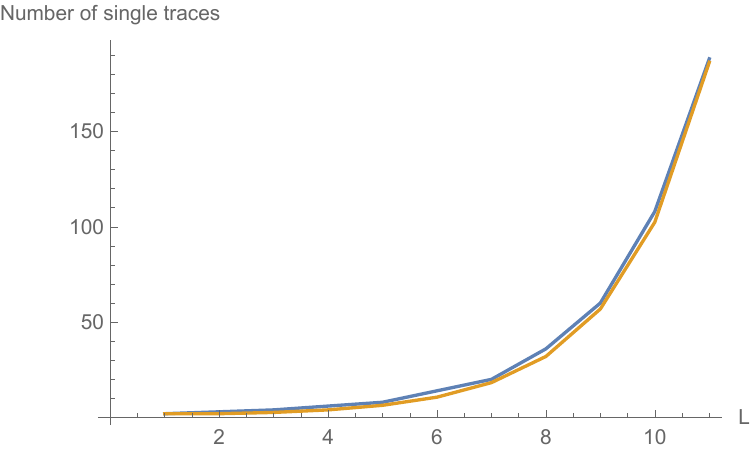}
\caption{Comparison of approximate vs exact counting for $N=2$. The exact counting curve lies slightly above the approximate counting curve.}
\label{CountNis2}
\end{center}
\end{figure}

To assess the accuracy of this estimate, we compare it to the exact count obtained via Pólya counting, as reviewed in Appendix \ref{PolyaCounting}. This comparison is illustrated in Figure \ref{CountNis2}, where we focus on the case $N=2$, since in this instance, the difference between the exact count and the approximate count is the most pronounced.

These numerical results confirm that $N_{\rm op}(L)$ is an excellent approximation for the number of single-trace operators of length $L$. In particular, the number of such operators grows exponentially with $L$. For a modest value of $N$, the total number of single trace operators of length $N$ which scales as $\sim e^N$, vastly exceeds the number of primary invariants ($=(d-1)N^2+1$). Thus, although all single-trace operators with $\leq N$ matrices are included as generating invariants, only a small fraction are actually primary invariants. 

To build some intuition, consider a two matrix model at $N=20$. Pólya counting predicts the total number of single-trace operators with $\leq N$ matrices is 111,321. The number of primary invariants is given by
\bea
M=N^2+1 = 401.
\eea
Thus at most only 0.36\% of the invariants with length $\le N$ are included as primary invariants and at least 99.64\% are included as secondary invariants. This demonstrates that extracting the primary invariants from the full set of gauge-invariant loops becomes increasingly selective as $N$ grows. This is in stark contrast to a naive cut off, motivated by the one matrix example, that would retain traces with length $\leq N$ as an independent set of finite $N$ generators.

\subsubsection{High Temperature Limit}

The multi matrix model partition functions we have computed all take the form of a polynomial divided by a denominator:
\bea
Z(x)=\frac{N(x)}{\prod_i (1-x^{n_i})^{m_i}}\label{pfcartoon}
\eea
with
\bea
\sum_i m_i = (d-1)N^2+1.
\eea
Since $x=e^{-\beta}$, the high-temperature limit $\beta\to 0$ corresponds to $x\to 1$. The polynomial $N(x)$ appearing in the numerator is a sum of positive terms, so it tends smoothly to a constant. Additionally, in this limit, we have
\bea
1-x^{n_i} = (1-x)(1+x+x^2+\cdots+x^{n_i-1})\to n_i (1-x).
\eea
Consequently, in the $x\to 1$ limit, the partition function simplifies to
\bea
Z(x)=\frac{k_{N}}{(1-x)^{(d-1)N^2+1}},
\eea
where $k_{N}$ is an $x$-independent constant that depends on $N$. Rewriting in terms of $\beta$, we use $(1-x) \approx \beta$ to obtain
\bea
\lim_{T\to\infty}Z(\beta) \simeq \frac{k_{N}}{(1-x)^{(d-1)N^2+1}} = k_{N}T^{(d-1)N^2+1}.
\eea
Making the reasonable assumption that the higher power of $x$ in the numerator of (\ref{pfcartoon}) scales as $N^2$, we should scale $T$ at least as $N^2$ at large $N$. The free energy is then given by
\bea
F &=& -T\log Z \,\,=\,\, -T\log k_{N} - \left((d-1)N^2+1\right)T\log T.
\eea
The entropy $S$ follows from
\bea
S=-\frac{\partial F}{\partial T},
\eea
which yields
\bea
S = \log k_{N} + \left((d-1)N^2+1\right)+ \left((d-1)N^2+1\right)\log T.
\eea
Thus, the entropy exhibits a characteristic growth as $N^2$. This is in perfect harmony with the analysis of \cite{troels}.

\subsubsection{The spectrum of secondary invariants}

Our two matrix model partition functions (\ref{pfcartoon}) take the form of a polynomial divided by a denominator, with a total of $N^2+1$ factors in the denominator. The numerator is a sum of positive powers of $x$, starting with 1 and with a final term that also has coefficient 1:
\bea
N(x) = 1+\cdots+x^K.
\eea
\begin{table}[h]
    \centering
    \begin{tabular}{|c|c|c|c|}
        \hline
        \textbf{N} & $\sum_i n_i m_i$ & $K$ & $\sum_i n_i m_i - 2 N^2$\\
        \hline
        \hline
        1 & 2     & 0    & 0\\ \hline
        2 & 8     & 0    & 0 \\ \hline
        3 & 24   & 6    & 6  \\ \hline
        4 & 56   & 24  & 24  \\ \hline
        5 & 104 & 54  & 54  \\ \hline
        6 & 181 & 109 & 109 \\ \hline
        7 & 278 & 180 & 180 \\ \hline
        \hline
    \end{tabular}
    \caption{Data extracted from the $d=2$ matrix partition functions.}
    \label{tab:example}
\end{table}
From the data in Table~\ref{tab:example}, it is evident that $K$ and $\sum_i n_i m_i$ are not independent quantities. The relationship between them is captured by the equation
\bea
K = \sum_i n_i m_i - 2N^2
\eea
derived using the temperature inversion symmetry discussed above.

The two-matrix model partition functions evaluated above allow a count of the number of primary and secondary invariants, as a function of $N$. This counting is summarized in Table \ref{tab:growth} below. The number of primary invariants is $N^2+1$, as already noted. The growth in the number of secondary invariants is clearly much more rapid than a power.
\begin{table}[h]
    \centering
    \begin{tabular}{|c|c|c|} % Three columns, left-aligned milena
        \hline
        $N$ & Primary Invariants & Non-trivial Secondary Invariants\\ 
        \hline\hline
        2 & 5  &  0 \\ \hline% done
        3 & 10 & 1 \\ \hline% done
        4 & 17 & 48 \\ \hline% done
        5 & 26 & 11,567 \\ \hline% done
        6 & 37 & 156,303,360 \\ \hline%done
        7 & 50 & 21,739,438,196,735 \\ \hline
        \hline
    \end{tabular}
    \caption{Growth in the number of invariants as $N$ increases.}
    \label{tab:growth}
\end{table}

Making the ansatz $N_{\rm Secondary}=e^{\alpha N^2-\beta N}$ and fitting to match the log of the number of secondary invariants, we find the result given in Figure \ref{fig:accident} below. The plot uses the values $\alpha=1.1$ and $\beta=3.3$. These results clearly suggest that
\bea
N_{\rm Secondary}=e^{c_1 N^2 - c_2 N}
\eea
where we could determine $c_1$ and $c_2$ more precisely with more data points. These two terms in the exponent are presumably the first in an infinite series that would include negative powers of $N$ and powers of $\log N$.
\begin{figure*}[h]
\includegraphics[width=0.75\linewidth]{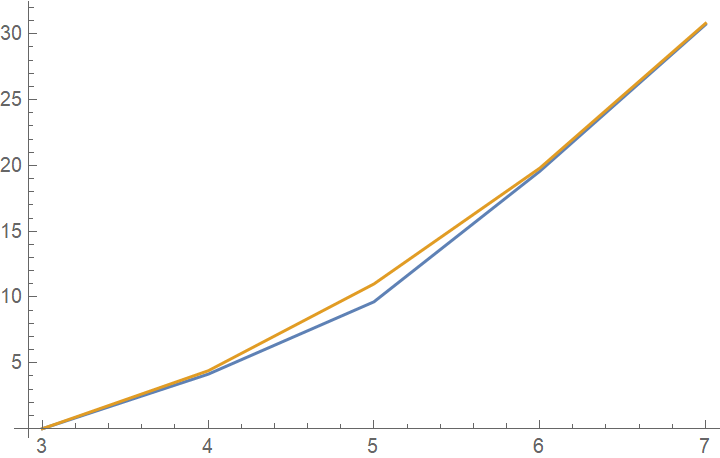}
\caption{The orange curve is $1.1 N^2 - 3.3 N$ while the blue curve is the log of the number of secondary invariants.} 
\label{fig:accident}
\end{figure*}

We have already seen that the denominator of the partition function dictates the high-temperature behaviour ($x\to 1$). However, the extremely rapid growth in the number of secondary invariants in the numerator suggests that, away from the high-temperature limit, these terms play a crucial role.

\section{Discussion}\label{discuss}

We have studied the structure of the space of gauge-invariant operators in a $d$-matrix model at finite $N$. For a single matrix, this space is the ring generated by single traces containing no more than $N$ matrices per trace. One might naturally conjecture that a similar result holds for multi-matrix models, with all single traces of length $\leq N$ forming a generating set. However, our results go significantly beyond and completely revise this naive expectation. 
As such they lead to substantial modifications of the commonly used  stringy exclusion principle \cite{Maldacena:1998bw}.

In summary the decompostion of Hilbert space goes as follows: at finite N we have identified a complete set of generating invariants, which fall into two distinct categories: \emph{primary} and \emph{secondary} invariants. Primary invariants are algebraically independent and generate the ring freely, while secondary invariants satisfy quadratic relations. Importantly, although all single-trace operators with $\leq N$ matrices appear among the generators, the generating set also includes invariants involving traces of more than $N$ matrices. Equally importantly, in no sense do traces of matrices with $\leq N$ matrices represents a complete set.

We have focused on free multi-matrix models, but we expect our results to extend naturally to interacting theories. Trace relations are purely kinematical constraints on gauge-invariant quantities and remain valid, unchanged, in the presence of interactions. As our algebraic analysis demonstrates, the interpretation of the partition functions depends solely on these trace relations. This includes key structural questions: whether a given set of invariants generates the full space of gauge-invariant operators, whether an invariant is quadratically reducible and what constraints exist among the invariants. Therefore, much of the structure of the gauge-invariant operator space at finite $N$ is largely insensitive to interactions.

At infinite $N$, there are no constraints among single-trace invariants, and the full space of gauge-invariant operators is freely generated by them,defining the extended collective field theory. All single-trace operators become primary invariants. As seen in our analysis, increasing $N$ promotes certain secondary invariants to primary status. In the $N\to\infty$ limit, all finite-length single-trace operators are primary, and the space of gauge-invariant operators is freely generated by their combinations.

As we have shown the growth in the number of secondary invariants ,at large N mirrors the expected scaling of black hole entropy. When generating the full space of gauge-invariant operators, primary invariants act freely -- they can appear with arbitrary multiplicity -- naturally giving rise to a Fock space structure. As such, they are best interpreted as perturbative degrees of freedom.

In contrast, each secondary invariant can appear only once and only linearly, resembling the behavior of a fixed background. In this sense, secondary invariants play the role of non-perturbative backgrounds that can be populated by perturbative excitations.\footnote{From this perspective, the Hironaka decomposition resembles the decomposition of a Fock space into a direct sum of superselection sectors.}

This then suggests a compelling interpretation: secondary invariants may correspond to black hole microstates. Furthermore, the promotion of secondary to primary invariants as $N$ increases can be viewed as a purely bosonic analogue of the fortuity mechanism. A natural framework to formalize these ideas is provided by the collective field theory representation\cite{Jevicki:1979mb,deMelloKoch:2002nq}. This connection, along with a more physical interpretation of our results, is explored further in \cite{ourpaper}.

\begin{center} 
{\bf Acknowledgements}
\end{center}
We would like to thank Sumit Das, Vishnu Jejjala and Costas Zoubos for discussions and Yiming Chen for useful correspondence. The work of RdMK is supported by a start up research fund of Huzhou University, a Zhejiang Province talent award and by a Changjiang Scholar award. The work of A.J. is supported by the U.S. Department of Energy under contract DE-SC0010010.

\appendix

\section{Counting Formulas}

In this Appendix we collect the counting formulas used in this paper, together with relevant background and references to the original literature.

\subsection{Counting Single Traces}\label{PolyaCounting}

Our primary invariants are single trace operators. For this reason, it is useful to be able to count the number of single trace operators constructed using a given number of $X,Y$ fields. This counting is easily performed using Polya theory - see \cite{Bianchi:2003wx}. For two matrices we define the single letter partition function
\bea
Z_1=x+y
\eea
and the single trace partition function
\bea
F(x,y)=\sum_{n}\sum_{n|d}{\varphi (d)\over n}Z_1 (x^d,y^d)^{n\over d}
\eea
The sum is over all positive integers $n$, and at each $n$ there is a second sum over $d$ which runs over the divisors of $n$, i.e. all the integers that can be divided into $n$ without remainder. $\varphi (d)$ is the Euler totient function. Euler's totient function counts the positive integers up to a given integer $n$ that are relatively prime to $n$. Two integers $a$ and $b$ are relatively prime if the only positive integer that divides both of them is $1$. $\varphi (9)=6$ because $1, 2, 4, 5, 7, 8$ are all relatively prime to $9$, while
$\varphi(12)=4$ because $1,5,7,11$ are relatively prime to $12$. The result for the single trace partition function is:
\bea
F(x,y)&=& (x+y)\cr
&+&\left(x^2+x y+y^2\right)\cr
&+&\left(x^3+x^2 y+x y^2+y^3\right)\cr
&+&\left(x^4+x^3 y+2 x^2 y^2+x y^3+y^4\right)\cr
&+&\left(x^5+x^4 y+2 x^3 y^2+2 x^2 y^3+x y^4+y^5\right)\cr
&+&\left(x^6+x^5 y+3 x^4 y^2+4 x^3 y^3+3 x^2 y^4+x y^5+y^6\right)\cr
&+&\left(x^7+x^6 y+3 x^5 y^2+5 x^4 y^3+5 x^3 y^4+3 x^2 y^5+x y^6+y^7\right)\cr
&+&\left(x^8+x^7 y+4 x^6 y^2+7 x^5 y^3+10 x^4 y^4+7 x^3 y^5+4 x^2 y^6+x y^7+y^8\right)\cr
&+&\left(x^9+x^8 y+4 x^7 y^2+10 x^6 y^3+14 x^5 y^4+14 x^4 y^5+10 x^3 y^6+4 x^2 y^7+x y^8+y^9\right)\cr
&+&\left(x^{10}+x^9 y+5 x^8 y^2+12 x^7 y^3+22 x^6 y^4+26 x^5 y^5+22 x^4 y^6+12 x^3 y^7+5 x^2 y^8
\right.\cr
&&\qquad\qquad\left. +x y^9+y^{10}\right)\cr
&+&\left(x^{11}+x^{10} y+5 x^9 y^2+15 x^8 y^3+30 x^7 y^4+42 x^6 y^5+42 x^5 y^6+30 x^4 y^7+15 x^3 y^8
\right.\cr
&&\qquad\qquad\left.+5 x^2 y^9+x y^{10}+y^{11}\right)\cr
&+&\left(x^{12}+x^{11} y+6 x^{10} y^2+19 x^9 y^3+43 x^8 y^4+66 x^7 y^5+80 x^6 y^6+66 x^5 y^7+43 x^4 y^8
\right.\cr
&&\qquad\qquad\left.+19 x^3 y^9+6 x^2 y^{10}+x y^{11}+y^{12}\right)\cr
&+& \left(x^{13}+x^{12} y+6 x^{11} y^2+22 x^{10} y^3+55 x^9 y^4+99 x^8 y^5+132 x^7 y^6+132 x^6 y^7
+99 x^5 y^8\right.\cr
&&\qquad\qquad\left. +55 x^4 y^9+22 x^3 y^{10}+6 x^2 y^{11}+x y^{12}+y^{13}\right)+\cdots
\nonumber
\eea

\subsection{Counting using Schur and Restricted Schur polynomials}\label{CountSandRS}

The Schur polynomials $\chi_R(Z)$ are a complete set of operators built out of a single $N\times N$  matrix field $Z$. Thus, to count all operators we simply need to count all Schur polynomials. At finite $N$ only operators $\chi_R(Z)$ labeled by Young diagrams $R$ with no more than $N$ rows are non-zero and are counted. The trace relations are the statement that $\chi_R(Z)=0$ for any Young diagram $R$ with more than $N$ rows. Each such equation gives an independent trace relation and together they give the complete set of trace relations. The partition function counting the complete set of operators is
\bea
Z(x)=\sum_{n=0}^\infty \sum_{\substack{R\vdash n\\ l(R)\le N}} x^{|R|}
\eea
where $|R|$ is the number of boxes in $R$ which is also the number of fields in the Schur polynomial, $R\vdash n$ means that $R$ is a partition of $n$, i.e. a Young diagram with $n$ boxes and $l(R)\le N$ means the number of rows in $R$ is less than or equal to $N$. Expanding $Z(x)$ in a power series the coefficient of $x^m$ counts how many gauge invariant operators can be constructed using $m$ $Z$ fields. Most of these are multi-trace operators. The number of independent trace relations $N_{Tr,M}$ generated using $M>N$ fields is given by
\bea
N_{Tr,M}=\sum_{\substack{R\vdash M\\ l(R)>N}}
\eea

The restricted Schur polynomials $\chi_{R,(r,s)\alpha\beta}(X,Y)$ are a complete set of operators built out of two $N\times N$ matrices $X,Y$ \cite{Bhattacharyya:2008rb,Bhattacharyya:2008xy}. for an operator constructed with $n$ $X$ fields and $m$ $Y$ fields, $R$ is a Young diagram with $n+m$ boxes, $r$ is a Young diagram with $n$ boxes and $s$ is a Young diagram with $m$ boxes. $\alpha,\beta$ are multiplicity labels distinguishing different copies of $(r,s)$ that arise after restricting the $S_{n+m}$ representation $R$ to the representation $(r,s)$ of the subgroup $S_n\times S_m$. $\alpha,\beta$ each run from 1 to $f_{rsR}$ where $f_{rsR}$ is the Littlewood-Richardson coefficient counting how many times $U(N)$ representation $R$ appears in the tensor product of $r$ and $s$. At finite $N$ only operators labelled by Young diagrams $R$ with no more than $N$ rows are non-zero and should be counted. We do not need to impose this restriction on $r$ and $s$ -- it is automatic that if $R$ has no more than than $N$ rows then any representation $(r,s)$ obtained from $R$ by restricting to a subgroup will only produce $r$'s and $s$'s with no more than $N$ rows. The trace relations are the statement that $\chi_{R,(r,s)\alpha\beta}(X,Y)=0$ for any Young diagram $R$ with more than $N$ rows. Each such equation gives an independent trace relation and together they give the complete set of trace relations. Thus, the partition function counting the complete set of operators is
\bea
Z(x,y)=\sum_{n=0}^\infty\sum_{m=0}^\infty \sum_{\substack{R\vdash n\\ l(R)\le N}}\sum_{r\vdash n}\sum_{s\vdash m} (f_{rsR})^2 x^{|r|}y^{|s|}\label{LRPF}
\eea
Expanding $Z(x,y)$ in a power series the coefficient of $x^n y^m$ counts how many gauge invariant operators can be constructed using $n$ $X$ and $m$ $Y$ fields. Most are multi-trace operators. The number of independent trace relations $N_{Tr,n,m}$ generated using $n$ $X$ and $m$ $Y$ fields ($m+n>N$ for a non-zero number) is given by
\bea
N_{Tr,n,m}=\sum_{\substack{R\vdash m+n\\ l(R)>N}}\sum_{r\vdash n}\sum_{s\vdash m}(f_{rsR})^2
\eea

For a model that involves $q$ matrices there is a natural generalization of the 2 matrix discussion. In this case the restricted Schur polynomials are $\chi_{R,(r_1,\cdots,r_q)\alpha\beta}(X_1,\cdots,X_q)$. For an operator constructed with $n_i$ $X_i$ fields $R$ is a Young diagram with $\sum_i n_i$ boxes and $r_i$ is a Young diagram with $n_i$ boxes. $\alpha,\beta$ are multiplicity labels which each run from 1 to $f_{r_1 r_2\cdots r_q R}$ where $f_{r_1 r_2\cdots r_q R}$ is the number of times $U(N)$ representation $R$ appears in the tensor product of $r_1\otimes r_2\otimes\cdots\otimes r_q$. At finite $N$ only operators labelled by Young diagrams $R$ with no more than $N$ rows are non-zero and should be counted. The trace relations are the statement that the restricted Schur polynomial vanishes for any Young diagram $R$ with more than $N$ rows. Each such equation gives an independent trace relation and together they give the complete set of trace relations. Thus, the partition function counting the complete set of operators is
\bea
Z(x_1,\cdots,x_q)=\sum_{n_1=0}^\infty\cdots \sum_{n_q=0}^\infty \sum_{\substack{R\vdash \sum_i n_i\\ l(R)\le N}}\sum_{r_1\vdash n_1}\cdots \sum_{r_q\vdash n_q} (f_{r_1 r_2\cdots r_qR})^2 \prod_{j=1}^qx_j^{|r_j|}
\label{LRPFMM}
\eea
and the number of independent trace relations generated using $n_i$ $X_i$ fields ($\sum_i n_i >N$ for a non-zero number) is given by
\bea
N_{Tr,n_1,n_2,\cdots,n_q}=\sum_{\substack{R\vdash \sum_in_i\\ l(R)>N}}\sum_{r_1\vdash n_1}\cdots \sum_{r_q\vdash n_q}(f_{r_1 r_2\cdots r_qR})^2
\eea

\subsection{Molien-Weyl Partition Functions}\label{MolienWeylPartitionFunctions}

As we explained in the last subsection we can count the total number of gauge invariant operators. Practically we are limited to about 30 operators in the trace. For a reasonable value of $N$ this can already take about a day to run on a laptop. In this subsection we will explain how to compute the same quantity using the Molien-Weyl formula. The Molien-Weyl formula was used for counting in \cite{Dolan:2007rq}. The benefit of the Molien-Weyl formula is that we can compute the exact partition function rather simply.

We count all gauge invariant composite operators constructed from bosonic fields with energies $E_i$ and in the adjoint of gauge group $U(N)$. The exact partition function is \cite{Sundborg:1999ue,Aharony:2003sx}
\bea
Z(\beta) = \sum_{n_1=0}^\infty x^{n_1E_1} \sum_{n_2=0}^\infty x^{n_2E_2} \cdots \times \#(n_1,n_2,...)
\eea
where $x=e^{-1/T}=e^{-\beta}$ and $\#(n_1,n_2,...)$ is the number of singlets in the tensor product $\sym^{n_1}_{adj}\otimes \sym^{n_2}_{adj}\otimes \cdots $. We must take the symmetric product because our fields are bosons. The number of singlets can be written as an integral over $U(N)$ of the product of the characters of the representations being tensored. Denote the Haar measure by $DU$. After rewriting we have
\bea
Z(\beta)&=&\int_{U(N)}\left[ DU\right] \,\, \prod_i \sum_{n_i=0}^\infty x^{n_iE_i} \chi_{\sym^{n_i}_{adj}}(U) 
\eea
It is simple to establish the character formula
\bea
\sum_{n=0}^\infty x^n\chi_{\sym^n_R}(U)&=&\exp{\sum_{m=1}^\infty\frac{1}{m} x^m \chi_R (U^m)}  
\eea
which holds for any representation $R$. For us $R$ is the adjoint representation, and we have 
\bea \label{eq:derivation_step4}
\mathcal{Z}(\beta)&=&\int_{U(N)}\left[ DU\right]\prod_i\exp{\sum_{m=1}^\infty \frac{1}{m}x^{mE_i}\chi_{adj}(U^m)}
\eea
The adjoint characters are $\chi_{adj}(U)={\rm Tr}U {\rm Tr}U^\dag$. In the fundamental of $U(N)$, the group elements are $N\times N$ matrices. Let $\varepsilon_j$ denote the $N$ eigenvalues of $U$. We have $|\varepsilon_j|=1$. In terms of these eigenvalues
\bea
\chi_{adj}(U^m)&=&{\rm Tr}U^m{\rm Tr}U^{\dag m}\,\,=\,\,\sum_{k,r=1}^N (\frac{\varepsilon_r}{\varepsilon_k})^m
\eea
Using this in (\ref{eq:derivation_step4}) we find
\bea
\mathcal{Z}(\beta)&=&\int_{U(N)}\!\left[ D U\right]\prod_{k,r=1}^N\prod_{i}\exp{ \sum_{m=1}^\infty \frac{1}{m} x^{mE_i} (\frac{\varepsilon_r}{\varepsilon_k})^m}\cr\cr
&=&\int_{U(N)}\left[ D U\right] \prod_{k,r=1}^N \frac{1}{\prod_i (1-x^{E_i}\frac{\varepsilon_r}{\varepsilon_k})}\cr
&=&\frac{1}{\prod_i (1-x^{E_i})^N}   \int_{U(N)}\left[ D U\right] \prod_{1\leq k<r \leq N} \frac{1}{\prod_i (1-x^{E_i}\frac{\varepsilon_r}{\varepsilon_k})(1-x^{E_i}\frac{\varepsilon_k}{\varepsilon_r})} \ .
\eea
To obtain the second line we did the sum over $m$ using $-\log(1-x) = \sum_{m=1}^\infty \frac{1}{m}x^m$. To obtain the last line, split the product into parts with $k=r$, $k<r$ and $k>r$. Only eigenvalues appear in the integrand. Performing the angular integrations, the problem reduces to an integral over the eigenvalues of the $U(N)$ group element $U$, i.e. we have an integral localized on the unit circle $|\varepsilon_j|=1$
\bea
\int_{U(N)}\left[ D U\right] \to \frac{1}{N! (2\pi i)^N} \oint \prod_{j=1}^N \frac{d \varepsilon_j}{\varepsilon_j} \ \Delta\bar{\Delta}
\eea
where we have introduced the Vandermonde determinants
\bea
 \Delta = \prod_{k<r} (\varepsilon_r-\varepsilon_k)\,\,=\,\,\sum_{\sigma\in S_N} \sgn(\sigma)\ \varepsilon_{\sigma(1)}^0\varepsilon_{\sigma(2)}^1 \cdots \varepsilon_{\sigma(N)}^{N-1}\qquad\bar{\Delta} = \prod_{k<r} (\varepsilon_r^{-1}-\varepsilon_k^{-1})\,.\label{vddet}
\eea
The partition function reduces to
\bea\label{eq:Molien_Weyl_alternative}
Z(x) = (Z_{N=1}(x))^N \frac{1}{N! (2\pi i)^N} \oint \prod_{j=1}^N \frac{d\varepsilon_j}{\varepsilon_j}\Delta\bar{\Delta}\prod_{1\leq k<r \leq N} \frac{1}{f_{k,r}}
\eea
where
\bea
Z_{N=1}(x) &=& \frac{1}{\prod_i (1-x^{E_i})}\qquad
f_{k,r}\,\,=\,\,\prod_i (1-x^{E_i}\frac{\varepsilon_r}{\varepsilon_k})(1-x^{E_i}\frac{\varepsilon_k}{\varepsilon_r})
\eea
To simplify this formula, use the symmetry under permutation of the $\varepsilon_i$. The measure $\prod_{j=1}^N \frac{d\varepsilon_j}{\varepsilon_j}$, $\Delta\bar{\Delta}$ and $\prod_{k<r}f_{k,r}(x)$ are invariant under permutation of $\varepsilon_i$. By permuting each term by $\sigma^{-1}$ in the sum appearing in the Vandermonde determinants (\ref{vddet}) we obtain
\bea
\Delta\bar{\Delta}&\to&N! \varepsilon_{1}^0\varepsilon_{2}^1 \cdots \varepsilon_{N}^{N-1} \bar{\Delta}\,\,=\,\,N! \prod_{k<r}(1-\frac{\varepsilon_r}{\varepsilon_k})
\eea
so that the partition function reduces to
\bea
Z(x)&=&\frac{(Z_{N=1}(x))^N}{(2\pi i)^N} \oint \prod_{j=1}^N \frac{d\varepsilon_j}{\varepsilon_j} \ \prod_{1\leq k<r \leq N} (1-\frac{\varepsilon_r}{\varepsilon_k}) \frac{1}{f_{k,r}}
\eea
Next, perform the change of variables $\varepsilon_j = t_1\cdots t_j$ so that
\bea
\prod_{1\leq k<r \leq N}\left(1 \pm x\frac{\varepsilon_r}{\varepsilon_k}\right)\left(1 \pm x\frac{\varepsilon_k}{\varepsilon_r}\right)&=&\prod_{2\leq k\leq r \leq N} (1 \pm x t_{k,r})\,(1 \pm xt_{k,r}^{-1})
\eea
and
\bea
\prod_{1\leq k<r\leq N}(1-\frac{\varepsilon_r}{\varepsilon_k})&=&\prod_{2\leq k\leq r\leq N} (1-t_{k,r})
\eea
where $t_{k,r}=t_k t_{k+1}\cdots t_r$. The Jacobian for the transformation is
\bea
J&=&\det\left[\frac{\partial\varepsilon_i}{\partial t_j}\right] = t_1^{N-1}t_2^{N-2}\cdots t_{N-1}^1 t_N^0 = \prod_{j=1}^N \frac{\varepsilon_j}{t_j} \ .
\eea
The partition function is now (we have relabeled $t_N\to t_{N-1}$, $t_{N-1}\to t_{N-2}, \dots$ $t_2\to t_1$ and $t_1\to t_N$)
\bea
Z(x)&=&\frac{(Z_{N=1}(x))^N}{(2\pi i)^N}\oint\prod_{j=1}^N\frac{dt_j}{t_j}
\prod_{1\leq k\leq r \leq N-1}\frac{1-t_{k,r}}{f_{k,r}}
\eea
and
\bea
Z_{N=1}(x) &=&\frac{1}{\prod_i (1-x^{E_i})} \qquad
f_{k,r}\,\,=\,\,\prod_i (1-x^{E_i}t_{k,r})(1-x^{E_i}t_{k,r}^{-1})
\eea
$t_N$ occurs once in the integrand as $1/t_N$. Integrating over $t_N$ then gives the final formula
\bea
Z(x) = \frac{(Z_{N=1}(x))^N}{(2\pi i)^{N-1}} \oint_{|t_1|=1} \frac{d t_1}{t_1} \cdots \oint_{|t_{N-1}|=1} \frac{d t_{N-1}}{t_{N-1}}   \prod_{1\leq k\leq r \leq N-1}  \frac{1-t_{k,r}}{f_{k,r}}\label{MWPF}
\eea
This is the formula we use to get the exact partition function. We only pick up residues inside the unit circle. Chemical potentials can again be included to get the graded partition function which keeps track of which species of field appears in the operator. Obviously this formula can be used for any number of matrices in the matrix model.

\section{Four Matrix Model at $N=2$}

In this Appendix we summarize some of the details that support the discussion of Section \ref{fourmatrixdiscusion}. 

\subsection{Quadratic Relations for Four Matrix Model at $N=2$}\label{solitonsN2M4}

In this subsection, we summarize the 10 relations that demonstrate the secondary invariants of the four-matrix model at $N=2$ are quadratically reducible. Specifically, these relations express any quadratic combination of the secondary invariants in terms of the complete set of invariants, but with at most a linear dependence in secondary invariants. Four secondary invariants give rise to 10 independent quadratic terms. The corresponding relations are given below.
\bea
s_1^2&=&m_{10} m_{2} s_{1}+\frac{m_{12}^2 m_{5}}{2}-\frac{1}{2} m_{12} m_{2} m_{3} m_{5}+\frac{1}{4} m_{2}^2 m_{5} m_{7}+\frac{1}{4} m_{3}^2 m_{5} m_{6}-\frac{1}{2}m_{3}^2 m_{9}^2+m_{3} m_{9} s_{1}\cr\cr
&&-\frac{1}{2} m_{1}^2 m_{12}^2+\frac{1}{2} m_{1}^2 m_{12} m_{2} m_{3}-\frac{1}{4} m_{1}^2 m_{2}^2 m_{3}^2+\frac{1}{4} m_{1}^2 m_{6} m_{7}+\frac{1}{2} m_{1} m_{10} m_{2}^2 m_{3}+m_{1} m_{12} s_1\cr\cr
&&-m_{1} m_{2} m_{3} s_{1}+\frac{1}{2} m_{1} m_{2} m_{3}^2 m_{9}
-m_{10} m_{12} m_{9}-\frac{1}{2} m_{1} m_{2} m_{7} m_{9}-\frac{1}{2}m_{10}^2 m_{2}^2+\frac{1}{2}m_{7} m_{9}^2\cr\cr
&&-\frac{1}{2} m_{1} m_{10} m_{3} m_{6}-\frac{1}{2}m_{5} m_{6} m_{7}+\frac{1}{2}m_{10}^2 m_{6}
\eea
\bea
s_2^2&=&m_{11} m_{2} s_{2}+\frac{1}{2}m_{13}^2 m_{5}-\frac{1}{2} m_{13} m_{2} m_{4} m_{5}+\frac{1}{4} m_{2}^2 m_{5} m_{8}+\frac{1}{4} m_{4}^2 m_{5} m_{6}-\frac{1}{2}m_{4}^2 m_{9}^2+m_{4} m_{9} s_{2}\cr\cr
&&-\frac{1}{2} m_{1}^2 m_{13}^2+\frac{1}{2} m_{1}^2 m_{13} m_{2} m_{4}-\frac{1}{4} m_{1}^2 m_{2}^2 m_{4}^2+\frac{1}{4} m_{1}^2 m_{6} m_{8}+\frac{1}{2} m_{1} m_{11} m_{2}^2 m_{4}+m_{1} m_{13} s_{2}\cr\cr
&&-m_{11} m_{13} m_{9}+\frac{1}{2} m_{1} m_{2} m_{4}^2 m_{9}-m_{1} m_{2} m_{4} s_{2}-\frac{1}{2} m_{1} m_{2} m_{8} m_{9}-\frac{1}{2}m_{11}^2 m_{2}^2+\frac{1}{2}m_{11}^2 m_{6}\cr\cr
&&-\frac{1}{2} m_{1} m_{11} m_{4} m_{6}-\frac{1}{2}m_{5} m_{6} m_{8}+\frac{1}{2}m_{8} m_{9}^2
\eea
\bea
s_3^2&=&m_{10} m_4 s_3-\frac{1}{2}m_{11}^2 m_3^2+\frac{1}{2}m_{11}^2 m_7 +m_{11} m_3 s_3+\frac{1}{2}m_{14}^2 m_5-\frac{1}{2} m_{14} m_3 m_4 m_5+\frac{1}{4} m_3^2 m_5 m_8\cr\cr
&&-\frac{1}{2}m_1^2 m_{14}^2+\frac{1}{2} m_1^2 m_{14} m_3 m_4-\frac{1}{4} m_{1}^2 m_3^2 m_4^2+\frac{1}{4} m_1^2 m_7 m_8+\frac{1}{2} m_1 m_{10} m_3 m_4^2+m_{1} m_{14} s_{3}\cr\cr
&&+\frac{1}{2} m_1 m_{11} m_3^2 m_{4}-\frac{1}{2} m_1 m_{10} m_3 m_8-m_1 m_3 m_4 s_{3}-\frac{1}{2}m_{10}^2 m_4^2+\frac{1}{2}m_{10}^2 m_8-m_{10} m_{11} m_{14}\cr\cr
&&-\frac{1}{2}m_1m_{11} m_4 m_7+\frac{1}{4}m_4^2 m_5m_7-\frac{1}{2}m_5m_7 m_8
\eea
\bea
s_4^2&=&m_{14} m_{2} s_{4}-\frac{1}{2} m_{14} m_{3} m_{4} m_{6}-\frac{1}{4} m_{2}^2 m_{3}^2 m_{4}^2+\frac{1}{4} m_{2}^2 m_{7} m_{8}-m_{2} m_{3} m_{4} s_{4}+\frac{1}{4} m_{3}^2 m_{6} m_{8}\cr\cr
&&-\frac{1}{2}m_{12}^2 m_{4}^2+\frac{1}{2}m_{12}^2 m_{8}-m_{12} m_{13} m_{14}+\frac{1}{2} m_{12} m_{2} m_{3} m_{4}^2-\frac{1}{2} m_{12} m_{2} m_{3} m_{8}+m_{12} m_{4} s_{4}\cr\cr
&&-\frac{1}{2}m_{13}^2 m_{3}^2+\frac{1}{2}m_{13}^2 m_{7}+\frac{1}{2} m_{13} m_{2} m_{3}^2 m_{4}-\frac{1}{2} m_{13} m_{2} m_{4} m_{7}+m_{13} m_{3} s_{4}-\frac{1}{2}m_{14}^2 m_2^2+\frac{1}{2}m_{14}^2 m_6\cr\cr
&&+\frac{1}{2} m_{14} m_{2}^2 m_{3} m_{4}+\frac{1}{4} m_{4}^2 m_{6} m_{7}-\frac{m_{6} m_{7} m_{8}}{2}
\eea
\bea
2s_1s_2&=&-m_{1}^2 m_{12} m_{13}+m_{1}^2 m_{12} m_{2} m_{4}+m_{1}^2 m_{13} m_{2} m_{3}+m_{1}^2 m_{14} m_{6}-\frac{1}{2} m_{1}^2 m_{2}^2 m_{3} m_{4}-m_{1}^2 m_{2} s_4\cr\cr
&&-\frac{1}{2} m_{1}^2 m_{3} m_{4} m_{6}-m_{1} m_{12} m_{4} m_{9}+2 m_{1} m_{12} s_{2}-m_{1} m_{14} m_{2} m_{9}+m_{1} m_{2}^2 s_{3}+m_{1} m_{2} m_{3} m_{4} m_{9}\cr\cr
&&-2 m_{1} m_{2} m_{3} s_{2}-m_{1} m_{6} s_{3}+m_{1} m_{9} s_{4}-m_{10} m_{11} m_{2}^2+m_{10} m_{11} m_{6}-m_{10} m_{13} m_{9}+2 m_{10} m_{2} s_{2}\cr\cr
&&-m_{11} m_{12} m_{9}+m_{11} m_{2} m_{3} m_{9}
+m_{12} m_{13} m_{5}-m_{12} m_{2} m_{4} m_{5}-m_{13} m_{2} m_{3} m_{5}-m_{14} m_{5} m_{6}\cr\cr
&&+m_{14} m_{9}^2+\frac{1}{2} m_{2}^2 m_{3} m_{4} m_{5}+m_{2} m_{5} s_{4}-m_{2} m_{9} s_{3}+\frac{1}{2} m_{3} m_{4} m_{5} m_{6}-m_{3} m_{4} m_{9}^2+m_{3} m_{9} s_{2}\cr\cr
&&+m_{4} m_{9} s_{1}
\eea
\bea
2s_1s_3&=&m_1^2 m_{12} m_{14}+m_1^2 m_{13} m_3^2-m_1^2 m_{13} m_7-\frac{1}{2} m_{1}^2 m_{2} m_{3}^2 m_{4}+\frac{1}{2} m_{1}^2 m_2 m_4 m_7-m_1 m_{10} m_{12} m_{4}\cr\cr
&&-m_1^2 m_3 s_4-m_{1} m_{10} m_{14} m_{2}+m_{1} m_{10} m_{2} m_{3} m_{4}+m_{1} m_{10} s_{4}-m_{1} m_{3}^2 s_{2}+m_1 m_7 s_2-m_{10}^2 m_{13}\cr\cr
&&+m_{10} m_{11} m_{12}-m_{10} m_{11} m_2 m_3+m_{10} m_{14} m_9+m_{10} m_2 s_3-m_{10} m_3 m_4 m_9+m_{10} m_3 s_2\cr\cr
&&+m_{10} m_4 s_1+m_{11} m_3^2 m_9-m_{11} m_7 m_9-m_{12} m_{14} m_5-m_{13} m_3^2 m_5+m_{13} m_5 m_7\cr\cr
&&+\frac{1}{2} m_2 m_3^2 m_4 m_5 -\frac{1}{2} m_2 m_4 m_5 m_7+m_3 m_5 s_4
\eea
\bea
2s_1s_4&=&m_1 m_{12} m_{14} m_2 - m_1 m_{12}^2 m_4+m_1 m_{12} m_2 m_3 m_{4}+m_{1} m_{12} s_{4}+m_{1} m_{13} m_{2} m_{3}^2-m_{1} m_{13} m_{2} m_{7}\cr\cr
&&+\frac{1}{2} m_{1} m_{2}^2 m_{4} m_{7}-\frac{1}{2} m_{1} m_{2}^2 m_{3}^2 m_{4}-2 m_{1} m_{2} m_{3} s_{4}-\frac{1}{2} m_{1} m_{3}^2 m_{4} m_{6}+\frac{1}{2} m_{1} m_{4} m_{6} m_{7}-m_{10} m_{12} m_{13}\cr\cr
&&-m_{10} m_{14} m_{2}^2+m_{10} m_{14} m_{6}+2 m_{10} m_{2} s_{4}+m_{11} m_{12}^2-m_{11} m_{12} m_{2} m_{3}+m_{11} m_{3}^2 m_{6}-m_{11} m_{6} m_{7}\cr\cr
&&-m_{12} m_{14} m_{9}-m_{12} m_{2} s_{3}-m_{12} m_{3} m_{4} m_{9}+m_{12} m_{3} s_{2}+m_{12} m_{4} s_{1}-m_{13} m_{3}^2 m_{9}+m_{2}^2 m_{3} s_{3}\cr\cr
&&+m_{13} m_7 m_9+m_2 m_3^2 m_4 m_9-m_2 m_3^2 s_2-m_2 m_4 m_7 m_9+m_2 m_7 s_2-m_3 m_6 s_3+2 m_3 m_9 s_4
\eea
\bea
2s_2s_3&=&\frac{1}{2} m_{1}^2 m_{12} m_{8}-m_{1}^2 m_{13} m_{14}+\frac{1}{2} m_{1}^2 m_{14} m_{2} m_{4}-\frac{1}{2} m_{1}^2 m_{2} m_{3} m_{4}^2+\frac{1}{2} m_{1}^2 m_{4} s_{4}+\frac{1}{2} m_{1} m_{10} m_{13} m_{4}\cr\cr
&&+\frac{1}{2} m_{1} m_{10} m_{2} m_{4}^2-\frac{1}{2} m_{1} m_{10} m_{2} m_{8}-\frac{3}{2} m_{1} m_{11} m_{12} m_{4}+\frac{3}{2} m_{1} m_{11} m_{2} m_{3} m_{4}+m_{1} m_{13} s_{3}\cr\cr
&&-\frac{1}{2} m_{1} m_{14} m_{4} m_{9}+m_{1} m_{14} s_{2}-\frac{3}{2} m_{1} m_{2} m_{4} s_{3}+m_{1} m_{3} m_{4}^2 m_{9}-\frac{1}{2} m_{1} m_{3} m_{4} s_{2}-\frac{1}{2} m_{1} m_{3} m_{8} m_{9}\cr\cr
&&-\frac{1}{2} m_{1} m_{4}^2 s_{1}-m_{10} m_{11} m_{13}-m_{10} m_{4}^2 m_{9}+m_{10} m_{8} m_{9}+m_{11}^2 m_{12}-m_{11}^2 m_{2} m_{3}-m_{11} m_{14} m_{9}\cr\cr
&&+m_{11} m_{2} s_{3}-m_{11} m_{3} m_{4} m_{9}+m_{11} m_{3} s_{2}+m_{11} m_{4} s_{1}+m_{12} m_{4}^2 m_{5}-m_{12} m_{5} m_{8}+m_{13} m_{14} m_{5}\cr\cr
&&-\frac{1}{2} m_{2} m_{3} m_{4}^2 m_{5}+\frac{1}{2} m_{2} m_{3} m_{5} m_{8}-m_{4} m_{5} s_{4}+2 m_{4} m_{9} s_{3}
\eea
\bea
2s_2s_4&=&-m_1 m_{12} m_{13} m_4+\frac{1}{2} m_1 m_{12} m_2 m_4^2+\frac{1}{2} m_1 m_{12} m_2 m_8-m_1 m_{13} m_{14} m_2+m_1 m_{13} m_2 m_3 m_4\cr\cr
&&+\frac{1}{2} m_{1} m_{14} m_{2}^2 m_{4}+\frac{1}{2} m_{1} m_{14} m_{4} m_{6}-\frac{1}{2} m_{1} m_{2}^2 m_{3} m_{4}^2-m_{1} m_{2} m_{4} s_{4}-\frac{1}{2} m_{1} m_{3} m_{6} m_{8}-m_{10} m_{13}^2\cr\cr
&&+m_{10} m_{13} m_{2} m_{4}-\frac{1}{2} m_{10} m_{2}^2 m_{8}-\frac{1}{2} m_{10} m_{4}^2 m_{6}+m_{10} m_{6} m_{8}+m_{11} m_{12} m_{13}-m_{11} m_{12} m_{2} m_{4}\cr\cr
&&-m_{11} m_{13} m_{2} m_{3}-m_{11} m_{14} m_{6}+\frac{1}{2} m_{11} m_{2}^2 m_{3} m_{4}+m_{11} m_{2} s_{4}+\frac{1}{2} m_{11} m_{3} m_{4} m_{6}+m_{12} m_{4} s_{2}\cr\cr
&&-m_{12} m_{8} m_{9}+m_{13} m_{14} m_{9}-m_{13} m_{3} m_{4} m_{9}+m_{13} m_{3} s_{2}-m_{14} m_{2} m_{4} m_{9}+m_{14} m_{2} s_{2}\cr\cr
&&+\frac{1}{2} m_2 m_3 m_4^2 m_9-m_2 m_3 m_4 s_2+\frac{1}{2} m_2 m_3 m_8 m_9+m_4 m_9 s_4+m_1 m_{13} s_4
\eea
\bea
2 s_3 s_4&=&m_1 m_{12} m_{14} m_4-\frac{1}{2} m_1 m_{12} m_3 m_8+m_1 m_{13} m_3^2 m_4-m_1 m_{13} m_4 m_7-m_1 m_{14}^2 m_2+m_1 m_{14} s_4\cr\cr
&&+\frac{1}{2} m_1 m_{14} m_2 m_3 m_4-\frac{1}{2} m_1 m_2 m_3^2 m_4^2+\frac{1}{2} m_1 m_2 m_4^2 m_7+\frac{1}{2}m_1m_2m_7m_8-\frac{3}{2}m_1m_3m_4 s_4\cr\cr
&&-m_{10} m_{12} m_4^2+m_{10} m_{12} m_8-m_{10} m_{13} m_{14}-\frac{1}{2} m_{10} m_{13} m_3 m_4+\frac{1}{2} m_{10} m_2 m_3 m_4^2-\frac{1}{2} m_{10} m_2 m_3 m_8\cr\cr
&&+2 m_{10} m_4 s_4-m_{11} m_{12} m_{14}+\frac{1}{2} m_{11} m_{12} m_3 m_4-m_{11} m_{13} m_3^2+m_{11} m_{13} m_7+\frac{1}{2} m_{11} m_2 m_3^2 m_4\cr\cr
&&-m_{11} m_2 m_4 m_7+m_{11} m_3 s_4+m_{13} m_3 s_3+m_{14}^2 m_9+m_{14} m_2 s_3-\frac{1}{2}m_{14}m_3m_4m_9-m_{14}m_4s_1\cr\cr
&&-\frac{1}{2}m_2m_3m_4 s_3-\frac{1}{2} m_3^2 m_4 s_2+\frac{1}{2} m_3^2 m_8 m_9+\frac{1}{2} m_3 m_4^2 s_{1}+m_4 m_7 s_2-m_7 m_8 m_9
\eea

\subsection{Constraints between generators of the ring}\label{constraintsN2M4}

The numerator of the partition function features four negative terms, which as we have discussed in Section \ref{fourmatrixdiscusion} correspond to constraints. The grading of each monomial specifies which matrices appear. Every term in the constraint corresponding to the monomial $w^n x^m y^p z^q$ is constructed using $n$ $W$'s, $m$ $X$'s, $p$ $Y$'s and $q$ $Z$'s.

The constraint corresponding to the monomial $x^2 w y z$ is given by
\bea
&&-\frac{1}{2} m_{1} m_{12} m_{2} m_{4}-\frac{1}{2}m_{1} m_{14} m_{6}+\frac{1}{2}m_{1} m_{2} s_{4}+\frac{1}{2} m_{1} m_{3} m_{4} m_{6}-\frac{1}{2}m_{10} m_{13} m_{2}+\frac{1}{2} m_{10} m_{2}^2 m_{4}\cr\cr
&&-\frac{1}{2}m_{10} m_{4} m_{6}+\frac{1}{2}m_{11} m_{12} m_{2}-\frac{1}{2}m_{11} m_{3} m_{6}+m_{12} m_{4} m_{9}-m_{12} s_{2}+m_{13} s_{1}+\frac{1}{2}m_{14} m_{2} m_{9}\cr\cr
&&-\frac{1}{2}m_{2}^2 s_{3}-\frac{1}{2} m_{2} m_{3} m_{4} m_{9}+\frac{1}{2}m_{2} m_{3} s_{2}-\frac{1}{2}m_{2} m_{4} s_{1}+m_{6} s_{3}-m_{9} s_{4}\,\,=\,\,0
\eea
The constraint corresponding to the monomial $x w y^2 z$ is given by
\bea
&&-m_{1} m_{12} m_{14}+\frac{1}{2} m_{1} m_{12} m_{3} m_{4}-\frac{1}{2} m_{1} m_{13} m_{3}^2+\frac{m_{1} m_{13} m_{7}}{2}+\frac{1}{2} m_{1} m_{14} m_{2} m_{3}-\frac{1}{2} m_{1} m_{2} m_{4} m_{7}\cr\cr
&&+\frac{1}{2}m_{1} m_{3} s_{4}+\frac{1}{2}m_{10} m_{13} m_{3}-m_{10} s_{4}-\frac{1}{2}m_{11} m_{12} m_{3}+\frac{1}{2}m_{11} m_{2} m_{7}+m_{12} s_{3}-\frac{1}{2}m_{14} m_{3} m_{9}\cr\cr
&&+m_{14} s_{1}-\frac{1}{2}m_{2} m_{3} s_{3}+\frac{1}{2}m_{3}^2 s_{2}-\frac{1}{2}m_{3} m_{4} s_{1}+\frac{1}{2}m_{4} m_{7} m_{9}-m_{7} s_{2}\,\,=\,\,0
\eea
The constraint corresponding to the monomial $x w y z^2$ is given by
\bea
&&-\frac{m_{1} m_{12} m_{8}}{2}-\frac{1}{2} m_{1} m_{14} m_{2} m_{4}+\frac{1}{2} m_{1} m_{2} m_{3} m_{8}+\frac{m_{1} m_{4} s_{4}}{2}-\frac{m_{10} m_{13} m_{4}}{2}+\frac{1}{2} m_{10} m_{2} m_{4}^2-\frac{m_{10} m_{2} m_{8}}{2}\cr\cr
&&+\frac{m_{11} m_{12} m_{4}}{2}+m_{11} m_{14} m_{2}-\frac{1}{2} m_{11} m_{2} m_{3} m_{4}-m_{11} s_{4}+m_{13} s_{3}+\frac{m_{14} m_{4} m_{9}}{2}-m_{14} s_{2}-\frac{m_{2} m_{4} s_{3}}{2}\cr\cr
&&+\frac{m_{3} m_{4} s_{2}}{2}-\frac{m_{3} m_{8} m_{9}}{2}-\frac{m_{4}^2 s_{1}}{2}+m_{8} s_{1}\,\,=\,\,0
\eea
The constraint corresponding to the monomial $x^2 w^2 y^2 z^2$ is given by
\bea
&&\frac{1}{2} m_{1}^2 m_{12} m_{13} m_{14}-\frac{1}{2} m_{1}^2 m_{12} m_{14} m_{2} m_{4}-\frac{1}{2} m_{1}^2 m_{13} m_{14} m_{2} m_{3}-\frac{1}{2} m_{1}^2 m_{14}^2 m_{6}+\frac{1}{4} m_{1}^2 m_{14} m_{2}^2 m_{3} m_{4}\cr\cr
&&+\frac{1}{2} m_{1}^2 m_{14} m_{2} s_{4}+\frac{1}{4} m_{1}^2 m_{14} m_{3} m_{4} m_{6}-m_{1} m_{12} m_{14} s_{2}+\frac{1}{4} m_{1} m_{12} m_{3} m_{4}^2 m_{9}-\frac{1}{4} m_{1} m_{13} m_{3}^2 m_{4} m_{9}\cr\cr
&&+\frac{1}{4} m_{1} m_{13} m_{4} m_{7} m_{9}+\frac{1}{2} m_{1} m_{14}^2 m_{2} m_{9}-\frac{1}{2} m_{1} m_{14} m_{2}^2 s_{3}-\frac{1}{4} m_{1} m_{14} m_{2} m_{3} m_{4} m_{9}+m_{1} m_{14} m_{2} m_{3} s_{2}\cr\cr
&&+\frac{1}{2} m_{1} m_{14} m_{6} s_{3}-\frac{1}{2} m_{1} m_{14} m_{9} s_{4}-\frac{1}{4} m_{1} m_{2} m_{4}^2 m_{7} m_{9}+\frac{1}{4} m_{1} m_{3} m_{4} m_{9} s_{4}+\frac{1}{2} m_{10} m_{11} m_{14} m_{2}^2\cr\cr
&&-\frac{1}{2} m_{10} m_{11} m_{14} m_{6}+\frac{1}{2} m_{10} m_{13} m_{14} m_{9}+\frac{1}{4} m_{10} m_{13} m_{3} m_{4} m_{9}-m_{10} m_{14} m_{2} s_{2}-\frac{1}{2} m_{10} m_{4} m_{9} s_{4}\cr\cr
&&+\frac{1}{2} m_{11} m_{12} m_{14} m_{9}-\frac{1}{4} m_{11} m_{12} m_{3} m_{4} m_{9}-\frac{1}{2} m_{11} m_{14} m_{2} m_{3} m_{9}+\frac{1}{4} m_{11} m_{2} m_{4} m_{7} m_{9}\cr\cr
&&+\frac{1}{2} m_{12} m_{14} m_{2} m_{4} m_{5}+\frac{1}{2} m_{12} m_{4} m_{9} s_{3}+\frac{1}{2} m_{13} m_{14} m_{2} m_{3} m_{5}+\frac{1}{2} m_{14}^2 m_{5} m_{6}-\frac{1}{2}m_{14}^2 m_{9}^2\cr\cr
&&-\frac{1}{4} m_{14} m_{2}^2 m_{3} m_{4} m_{5}-\frac{1}{2} m_{14} m_{2} m_{5} s_{4}+\frac{1}{2} m_{14} m_{2} m_{9} s_{3}-\frac{1}{4} m_{14} m_{3} m_{4} m_{5} m_{6}+\frac{1}{4} m_{14} m_{3} m_{4} m_{9}^2\cr\cr
&&-\frac{1}{2} m_{14} m_{3} m_{9} s_{2}+m_{14} s_{1} s_{2}-\frac{1}{4} m_{2} m_{3} m_{4} m_{9} s_{3}+\frac{1}{4} m_{3}^2 m_{4} m_{9} s_{2}-\frac{1}{4} m_{3} m_{4}^2 m_{9} s_{1}+\frac{1}{4} m_{4}^2 m_{7} m_{9}^2\cr\cr
&&-\frac{1}{2} m_{4} m_{7} m_{9} s_{2}-\frac{1}{2} m_{12} m_{13} m_{14} m_{5}\,\,=\,\,0
\eea

\section{Two Matrix Model at $N=3$}

In this Appendix we provide details used in the algebraic analysis of Section \ref{2matn3discuss}. The trace relations quoted in the next subsection are used to prove that extra generators must be introduced: after the trace relations are used, there are still some invariants that are not determined. The relations of Section \ref{solrel} establish that these generators are quadratically reducible, i.e. they are secondary invariants.

\subsection{Trace Relations}\label{TrRelN3}

The trace relations take the form $T_3(A,B,C,D)=0$ for any words $A$, $B$, $C$, and $D$, where $T_3(A,B,C,D)$ is given by
\begin{align}
T_3(A,B,C,D) &= \Tr(A)\Tr(B)\Tr(C)\Tr(D) \nonumber \\
&\quad - \Tr(AB)\Tr(C)\Tr(D) - \Tr(AC)\Tr(B)\Tr(D) - \Tr(AD)\Tr(B)\Tr(C) \nonumber \\
&\quad - \Tr(A)\Tr(BC)\Tr(D) - \Tr(A)\Tr(BD)\Tr(C) - \Tr(A)\Tr(B)\Tr(CD) \nonumber \\
&\quad + \Tr(AB)\Tr(CD) + \Tr(AC)\Tr(BD) + \Tr(AD)\Tr(BC) \nonumber \\
&\quad + \Tr(ABC)\Tr(D) + \Tr(ACB)\Tr(D) + \Tr(ABD)\Tr(C) + \Tr(ADB)\Tr(C) \nonumber \\
&\quad + \Tr(ACD)\Tr(B) + \Tr(ADC)\Tr(B) + \Tr(A)\Tr(BCD) + \Tr(A)\Tr(BDC) \nonumber \\
&\quad - \Tr(ABCD) - \Tr(ABDC) - \Tr(ACBD) \nonumber \\
&\quad - \Tr(ACDB) - \Tr(ADBC) - \Tr(ADCB).
\end{align}

\noindent
{\bf Degree 4:} There are six single trace operators that can be constructed using four matrices. There are only five independent trace relations among them. Using the trace relations, we find  
\bea
\Tr(X^4) &=& \frac{1}{6} \left(-6 m_{1}^2 m_{3} + m_{1}^4 + 8 m_{1} m_{6} + 3 m_{3}^2 \right),\cr\cr
\Tr(Y^4) &=& \frac{1}{6} \left(-6 m_{2}^2 m_{5} + m_{2}^4 + 8 m_{2} m_{9} + 3 m_{5}^2 \right),\cr\cr
\Tr(X^3Y) &=& \frac{1}{6} \left(m_{1}^3 m_{2} - 3 m_{1}^2 m_{4} - 3 m_{1} m_{2} m_{3} + 6 m_{1} m_{7} + 2 m_{2} m_{6} + 3 m_{3} m_{4} \right),\cr\cr
\Tr(XY^3) &=& \frac{1}{6} \left(m_{1} m_{2}^3 - 3 m_{1} m_{2} m_{5} + 2 m_{1} m_{9} - 3 m_{2}^2 m_{4} + 6 m_{2} m_{8} + 3 m_{4} m_{5} \right),\cr\cr
\Tr(XYXY) &=& \frac{1}{2} \left(m_{1}^2 m_{2}^2 - m_{1}^2 m_{5} - 4 m_{1} m_{2} m_{4} + 4 m_{1} m_{8} - 4 m_{10} - m_{2}^2 m_{3} + 4 m_{2} m_{7}\right.\cr\cr
&&\qquad\qquad\left. + m_{3} m_{5} + 2 m_{4}^2 \right).
\eea
The operator $\Tr(X^2Y^2)$ is not determined by these trace relations, so it must be included in the set of generating invariants.

\noindent
{\bf Degree 5:} There are eight invariants that are not included among the generating invariants. Using the trace relations, we can express them all in terms of the primary invariants.
\bea
\Tr(X^5)&=& \frac{1}{6} \left(-5 m_{1}^3 m_{3}+5 m_{1}^2 m_{6}+m_{1}^5+5 m_{3} m_{6}\right)\cr\cr
\Tr(Y^5)&=& \frac{1}{6} \left(-5 m_{2}^3 m_{5}+5 m_{2}^2 m_{9}+m_{2}^5+5 m_{5} m_{9}\right)\cr\cr
\Tr(X^4Y)&=& \frac{1}{6} \left(-3 m_{1}^2 m_{2} m_{3}+m_{1}^4 m_{2}-2 m_{1}^3 m_{4}+3 m_{1}^2 m_{7}+2 m_{1} m_{2} m_{6}+3 m_{3} m_{7}+2 m_{4} m_{6}\right)\cr\cr
\Tr(XY^4)&=& \frac{1}{6} \left(-3 m_{1} m_{2}^2 m_{5}+m_{1} m_{2}^4+2 m_{1} m_{2} m_{9}-2 m_{2}^3 m_{4}+3 m_{2}^2 m_{8}+2 m_{4} m_{9}+3 m_{5} m_{8}\right)\cr\cr
\Tr(X^3Y^2)&=& \frac{1}{6} \left(m_{1}^3 m_{5}-3 m_{1}^2 m_{8}+6 m_{1} m_{10}-3 m_{1} m_{3} m_{5}+3 m_{3} m_{8}+2 m_{5} m_{6}\right)\cr\cr
\Tr(X^2Y^3)&=& \frac{1}{6} \left(6 m_{10} m_{2}+m_{2}^3 m_{3}-3 m_{2}^2 m_{7}-3 m_{2} m_{3} m_{5}+2 m_{3} m_{9}+3 m_{5} m_{7}\right)\cr\cr
\Tr(X^2YXY)&=& \frac{1}{6} \left(2 m_{1}^3 m_{2}^2-6 m_{1}^2 m_{2} m_{4}-2 m_{1}^3 m_{5}+6 m_{1}^2 m_{8}-6 m_{1} m_{10}-3 m_{1} m_{2}^2 m_{3}+6 m_{1} m_{2} m_{7}\right.\cr\cr
&&\left.+3 m_{1} m_{3} m_{5}+m_{2}^2 m_{6}+6 m_{4} m_{7}-m_{5} m_{6}\right)\cr\cr
\Tr(XY^2XY)&=& \frac{1}{6} \left(2 m_{1}^2 m_{2}^3-3 m_{1}^2 m_{2} m_{5}+m_{1}^2 m_{9}-6 m_{1} m_{2}^2 m_{4}+6 m_{1} m_{2} m_{8}-6 m_{10} m_{2}-2 m_{2}^3 m_{3}\right.\cr\cr
&&\left.+6 m_{2}^2 m_{7}+3 m_{2} m_{3} m_{5}-m_{3} m_{9}+6 m_{4} m_{8}\right)
\eea

\noindent
{\bf Degree 6:} There are eight invariants that are not included among the generating invariants. Even after using trace relations, we find a single invariant that is not determined. Thus, one more invariant must be added to the generating set.
\bea
\Tr(X^6)&=& \frac{1}{12} \left(-9 m_{1}^2 m_{3}^2-3 m_{1}^4 m_{3}+4 m_{1}^3 m_{6}+m_{1}^6+12 m_{1} m_{3} m_{6}+3 m_{3}^3+4 m_{6}^2\right)\cr\cr
\Tr(Y^6)&=& \frac{1}{12} \left(-9 m_{2}^2 m_{5}^2-3 m_{2}^4 m_{5}+4 m_{2}^3 m_{9}+m_{2}^6+12 m_{2} m_{5} m_{9}+3 m_{5}^3+4 m_{9}^2\right)\cr\cr
\Tr(X^5Y)&=& \frac{1}{12} \left(-2 m_{1}^3 m_{2} m_{3}+2 m_{1}^2 m_{2} m_{6}+m_{1}^5 m_{2}-6 m_{1}^2 m_{3} m_{4}-m_{1}^4 m_{4}+2 m_{1}^3 m_{7}\right.\cr\cr
&&\left.-3 m_{1} m_{2} m_{3}^2+6 m_{1} m_{3} m_{7}+4 m_{1} m_{4} m_{6}+2 m_{2} m_{3} m_{6}+3 m_{3}^2 m_{4}+4 m_{6} m_{7}\right)\cr\cr
\Tr(XY^5)&=& \frac{1}{24} \left(-6 m_{1} m_{2}^3 m_{5}+6 m_{1} m_{2}^2 m_{5}+6 m_{1} m_{2}^2 m_{9}+3 m_{1} m_{2}^5-2 m_{1} m_{2}^4-9 m_{1} m_{2} m_{5}^2\right.\cr\cr
&&\left.-4 m_{1} m_{2} m_{9}+6 m_{1} m_{5} m_{9}-18 m_{2}^2 m_{4} m_{5}-3 m_{2}^4 m_{4}+4 m_{2}^3 m_{4}+6 m_{2}^3 m_{8}-6 m_{2}^2 m_{8}\right.\cr\cr
&&\left.+12 m_{2} m_{4} m_{9}+18 m_{2} m_{5} m_{8}+9 m_{4} m_{5}^2-4 m_{4} m_{9}-6 m_{5} m_{8}+12 m_{8} m_{9}\right)\cr\cr
\Tr(X^4Y^2)&=& \frac{1}{6} \left(3 m_{1}^2 m_{10}-3 m_{1}^2 m_{3} m_{5}+m_{1}^4 m_{5}-2 m_{1}^3 m_{8}+2 m_{1} m_{5} m_{6}+3 m_{10} m_{3}+2 m_{6} m_{8}\right)\cr\cr
\Tr(X^3YXY)&=& \frac{1}{12} \left(m_{1}^4 m_{2}^2-6 m_{1}^2 m_{4}^2-m_{1}^4 m_{5}+2 m_{1}^3 m_{8}+2 m_{1} m_{2}^2 m_{6}-12 m_{1} m_{2} m_{3} m_{4}+6 m_{1} m_{3} m_{8}\right.\cr\cr
&&\left.+12 m_{1} m_{4} m_{7}-2 m_{1} m_{5} m_{6}-12 m_{10} m_{3}-3 m_{2}^2 m_{3}^2+12 m_{2} m_{3} m_{7}+3 m_{3}^2 m_{5}+6 m_{3} m_{4}^2\right.\cr\cr
&&\left.+4 m_{6} m_{8}\right)\cr\cr
\Tr(X^2YX^2Y)&=& \frac{1}{12} \left(-12 m_{1}^2 m_{10}-6 m_{1}^2 m_{2}^2 m_{3}+3 m_{1}^4 m_{2}^2-8 m_{1}^3 m_{2} m_{4}+12 m_{1}^2 m_{2} m_{7}+6 m_{1}^2 m_{3} m_{5}\right.\cr\cr
&&\left.-3 m_{1}^4 m_{5}+8 m_{1}^3 m_{8}+12 m_{10} m_{3}+3 m_{2}^2 m_{3}^2-12 m_{2} m_{3} m_{7}+8 m_{2} m_{4} m_{6}-3 m_{3}^2 m_{5}\right.\cr\cr
&&\left.-8 m_{6} m_{8}+12 m_{7}^2\right)\cr\cr
\Tr(X^2Y^4)&=& \frac{1}{6} \left(3 m_{10} m_{2}^2+3 m_{10} m_{5}-3 m_{2}^2 m_{3} m_{5}+m_{2}^4 m_{3}-2 m_{2}^3 m_{7}+2 m_{2} m_{3} m_{9}+2 m_{7} m_{9}\right)\cr\cr
\Tr(XY^3XY)&=& \frac{1}{12} \left(m_{1}^2 m_{2}^4+2 m_{1}^2 m_{2} m_{9}-3 m_{1}^2 m_{5}^2-12 m_{1} m_{2} m_{4} m_{5}+12 m_{1} m_{5} m_{8}-12 m_{10} m_{5}\right.\cr\cr
&&\left.-m_{2}^4 m_{3}-6 m_{2}^2 m_{4}^2+2 m_{2}^3 m_{7}-2 m_{2} m_{3} m_{9}+12 m_{2} m_{4} m_{8}+6 m_{2} m_{5} m_{7}+3 m_{3} m_{5}^2\right.\cr\cr
&&\left.+6 m_{4}^2 m_{5}+4 m_{7} m_{9}\right)\cr\cr
\Tr(XY^2XY^2)&=& \frac{1}{12} \left(-6 m_{1}^2 m_{2}^2 m_{5}+3 m_{1}^2 m_{2}^4+3 m_{1}^2 m_{5}^2-8 m_{1} m_{2}^3 m_{4}+12 m_{1} m_{2}^2 m_{8}+8 m_{1} m_{4} m_{9}\right.\cr\cr
&&\left.-12 m_{1} m_{5} m_{8}-12 m_{10} m_{2}^2+12 m_{10} m_{5}+6 m_{2}^2 m_{3} m_{5}-3 m_{2}^4 m_{3}+8 m_{2}^3 m_{7}-3 m_{3} m_{5}^2\right.\cr\cr
&&\left.-8 m_{7} m_{9}+12 m_{8}^2\right)\cr\cr
\Tr(X^3Y^3)&=& \frac{1}{12} \left(3 m_{1}^2 m_{2}^2 m_{4}+m_{1}^3 \left(-m_{2}^3\right)+3 m_{1}^3 m_{2} m_{5}-6 m_{1}^2 m_{2} m_{8}-3 m_{1}^2 m_{4} m_{5}+12 m_{1} m_{10} m_{2}\right.\cr\cr
&&\left.+3 m_{1} m_{2}^3 m_{3}-6 m_{1} m_{2}^2 m_{7}-9 m_{1} m_{2} m_{3} m_{5}+6 m_{1} m_{5} m_{7}-3 m_{2}^2 m_{3} m_{4}+6 m_{2} m_{3} m_{8}\right.\cr\cr
&&\left.+3 m_{3} m_{4} m_{5}+4 m_{6} m_{9}\right)\cr\cr
\Tr(X^2Y^2XY)&=& \frac{1}{3} \left(-3 \Tr(X^2YXY^2)-3 m_{1}^2 m_{2}^2 m_{4}+m_{1}^3 m_{2}^3-m_{1}^3 m_{2} m_{5}+3 m_{1}^2 m_{2} m_{8}-3 m_{1} m_{10} m_{2}\right.\cr\cr
&&\left.-m_{1} m_{2}^3 m_{3}+3 m_{1} m_{2}^2 m_{7}+m_{1} m_{3} m_{9}+3 m_{10} m_{4}+m_{2} m_{5} m_{6}-m_{6} m_{9}+3 m_{7} m_{8}\right)\cr\cr
\Tr(XYXYXY)&=& \frac{1}{12} \left(9 m_{1}^2 m_{2}^2 m_{4}+m_{1}^3 m_{2}^3-3 m_{1}^3 m_{2} m_{5}-9 m_{1}^2 m_{4} m_{5}+2 m_{1}^3 m_{9}-3 m_{1} m_{2}^3 m_{3}\right.\cr\cr
&&\left.+9 m_{1} m_{2} m_{3} m_{5}-36 m_{1} m_{2} m_{4}^2-6 m_{1} m_{3} m_{9}+36 m_{1} m_{4} m_{8}-36 m_{10} m_{4}-9 m_{2}^2 m_{3} m_{4}\right.\cr\cr
&&\left.+2 m_{2}^3 m_{6}+36 m_{2} m_{4} m_{7}-6 m_{2} m_{5} m_{6}+9 m_{3} m_{4} m_{5}+12 m_{4}^3+4 m_{6} m_{9}\right)
\eea
Thus, the operator $\Tr(X^2YXY^2)$ is not determined by the trace relations and so must be included as an extra invariant in our generating set. In the next subsection we prove that this invariant is quadratically reducible.

\subsection{Coefficients of the secondary invariant relation}\label{solrel}

The additional generator introduced above satisfies the quadratic relation  
\bea
s_1^2 + p_1(m_i) s_1 + p_2(m_i) = 0,
\eea
demonstrating that it is quadratically reducible. While the explicit expressions for the coefficients in this relation are too complicated to provide direct insight, we include them here for completeness. The existence of this constraint is crucial for establishing our interpretation of the partition function. Moreover, given the constraint, its validity can be easily verified.
\bea
p_1(m_i)&=&m_1^2 m_2^2 m_4-\frac{m_1^3 m_2^3}{3}+\frac{1}{3} m_1^3 m_2 m_5-m_1^2 m_2 m_8+m_1 m_{10} m_2+\frac{1}{3} m_1 m_2^3 m_3-m_1 m_2^2 m_7\cr\cr
&&-\frac{m_1 m_3 m_9}{3}-m_{10} m_4-\frac{m_2 m_5 m_6}{3}+\frac{m_6 m_9}{3}-m_7 m_8
\eea

\bea
p_2(m_i)&=&\frac{5 m_2^6 m_1^6}{216}+\frac{m_5^3 m_1^6}{144}+\frac{m_9^2 m_1^6}{216}-\frac{5}{144} m_2^4 m_5 m_1^6+\frac{1}{8} m_2 m_4 m_5^2 m_1^5-\frac{1}{8} m_2^5 m_4 m_1^5\cr\cr
&&+\frac{1}{36} m_2^3 m_4 m_5 m_1^5+\frac{1}{9} m_2^4 m_8 m_1^5-\frac{1}{12} m_5^2 m_8 m_1^5-\frac{1}{36} m_4 m_5 m_9 m_1^5-\frac{1}{36} m_2 m_8 m_9 m_1^5\cr\cr
&&-\frac{1}{12} m_{10} m_2^4 m_1^4-\frac{1}{48} m_3 m_5^3 m_1^4+\frac{19}{144} m_2^4 m_4^2 m_1^4-\frac{1}{16} m_4^2 m_5^2 m_1^4+\frac{1}{6} m_{10} m_5^2 m_1^4\cr\cr
&&+\frac{5}{48} m_2^2 m_3 m_5^2 m_1^4+\frac{1}{24} m_2^2 m_8^2 m_1^4+\frac{7}{24} m_5 m_8^2 m_1^4-\frac{1}{24} m_3 m_9^2 m_1^4-\frac{5}{144} m_2^6 m_3 m_1^4\cr\cr
&&-\frac{1}{12} m_{10} m_2^2 m_5 m_1^4+\frac{5}{24} m_2^2 m_4^2 m_5 m_1^4-\frac{5}{144} m_2^4 m_3 m_5 m_1^4+\frac{1}{9} m_2^5 m_7 m_1^4-\frac{1}{6} m_2 m_5^2 m_7 m_1^4\cr\cr
&&-\frac{1}{6} m_2^3 m_4 m_8 m_1^4-\frac{7}{12} m_2 m_4 m_5 m_8 m_1^4+\frac{1}{18} m_2 m_4^2 m_9 m_1^4+\frac{1}{18} m_2^3 m_3 m_9 m_1^4\cr\cr
&&-\frac{1}{36} m_2 m_3 m_5 m_9 m_1^4+\frac{1}{18} m_2^2 m_7 m_9 m_1^4+\frac{1}{12} m_4 m_8 m_9 m_1^4+\frac{1}{9} m_2^3 m_4^3 m_1^3-\frac{5}{12} m_2 m_3 m_4 m_5^2 m_1^3\cr\cr
&&+\frac{2}{3} m_2 m_4 m_8^2 m_1^3+\frac{1}{27} m_6 m_9^2 m_1^3+\frac{1}{6} m_{10} m_2^3 m_4 m_1^3+\frac{1}{36} m_2^5 m_3 m_4 m_1^3+\frac{2}{3} m_{10} m_2 m_4 m_5 m_1^3\cr\cr
&&+\frac{1}{2} m_2^3 m_3 m_4 m_5 m_1^3-\frac{1}{18} m_5^3 m_6 m_1^3+\frac{1}{18} m_2^4 m_5 m_6 m_1^3+\frac{1}{6} m_4 m_5^2 m_7 m_1^3-\frac{1}{6} m_2^4 m_4 m_7 m_1^3\cr\cr
&&-\frac{1}{2} m_2^2 m_4 m_5 m_7 m_1^3+\frac{1}{6} m_{10} m_2^2 m_8 m_1^3-\frac{1}{2} m_2^2 m_4^2 m_8 m_1^3+\frac{1}{6} m_3 m_5^2 m_8 m_1^3+\frac{1}{6} m_4^2 m_5 m_8 m_1^3\cr\cr
&&-m_{10} m_5 m_8 m_1^3-\frac{1}{2} m_2^2 m_3 m_5 m_8 m_1^3+\frac{1}{6} m_2^3 m_7 m_8 m_1^3+\frac{2}{3} m_2 m_5 m_7 m_8 m_1^3-\frac{1}{9} m_4^3 m_9 m_1^3\cr\cr
&&-\frac{5}{18} m_2^2 m_3 m_4 m_9 m_1^3+\frac{1}{6} m_3 m_4 m_5 m_9 m_1^3-\frac{5}{54} m_2^3 m_6 m_9 m_1^3+\frac{1}{18} m_2 m_5 m_6 m_9 m_1^3\cr\cr
&&-\frac{1}{3} m_2 m_4 m_7 m_9 m_1^3+\frac{1}{3} m_2 m_3 m_8 m_9 m_1^3-\frac{m_8^3 m_1^3}{3}+\frac{1}{16} m_3^2 m_5^3 m_1^2-\frac{1}{4} m_{10}^2 m_2^2 m_1^2\cr\cr
&&+\frac{1}{4} m_{10} m_2^2 m_4^2 m_1^2+\frac{5}{24} m_2^4 m_3 m_4^2 m_1^2-\frac{1}{8} m_2^2 m_3^2 m_5^2 m_1^2+\frac{1}{8} m_3 m_4^2 m_5^2 m_1^2-\frac{1}{2} m_{10} m_3 m_5^2 m_1^2\cr\cr
&&+\frac{1}{24} m_2^4 m_7^2 m_1^2-\frac{1}{8} m_5^2 m_7^2 m_1^2+\frac{1}{4} m_2^2 m_5 m_7^2 m_1^2+\frac{3}{2} m_{10} m_8^2 m_1^2+\frac{1}{4} m_2^2 m_3 m_8^2 m_1^2-\frac{1}{4} m_3 m_5 m_8^2 m_1^2\cr\cr
&&-\frac{1}{2} m_2 m_7 m_8^2 m_1^2+\frac{1}{8} m_3^2 m_9^2 m_1^2-\frac{1}{12} m_{10} m_2^4 m_3 m_1^2+\frac{3}{4} m_{10}^2 m_5 m_1^2+\frac{5}{48} m_2^4 m_3^2 m_5 m_1^2\cr\cr
&&-\frac{1}{4} m_{10} m_4^2 m_5 m_1^2-\frac{1}{2} m_2^2 m_3 m_4^2 m_5 m_1^2+\frac{3}{4} m_{10} m_2^2 m_3 m_5 m_1^2-\frac{5}{18} m_2^3 m_4 m_5 m_6 m_1^2\cr\cr
&&+\frac{1}{6} m_{10} m_2^3 m_7 m_1^2-\frac{1}{2} m_2^3 m_4^2 m_7 m_1^2+\frac{1}{2} m_2 m_3 m_5^2 m_7 m_1^2-m_{10} m_2 m_5 m_7 m_1^2-\frac{1}{2} m_2^3 m_3 m_5 m_7 m_1^2\cr\cr
&&-\frac{3}{2} m_{10} m_2 m_4 m_8 m_1^2-\frac{1}{2} m_2^3 m_3 m_4 m_8 m_1^2+m_2 m_3 m_4 m_5 m_8 m_1^2+\frac{1}{18} m_2^4 m_6 m_8 m_1^2\cr\cr
&&+\frac{1}{3} m_5^2 m_6 m_8 m_1^2+m_2^2 m_4 m_7 m_8 m_1^2-\frac{1}{2} m_4 m_5 m_7 m_8 m_1^2+\frac{1}{6} m_2 m_3 m_4^2 m_9 m_1^2+\frac{1}{3} m_2 m_7^2 m_9 m_1^2\cr\cr
&&-\frac{1}{6} m_{10} m_2 m_3 m_9 m_1^2-\frac{1}{6} m_2 m_3^2 m_5 m_9 m_1^2+\frac{1}{2} m_2^2 m_4 m_6 m_9 m_1^2-\frac{2}{9} m_4 m_5 m_6 m_9 m_1^2\cr\cr
&&+\frac{1}{2} m_4^2 m_7 m_9 m_1^2-\frac{1}{6} m_{10} m_7 m_9 m_1^2-\frac{1}{2} m_3 m_4 m_8 m_9 m_1^2-\frac{7}{18} m_2 m_6 m_8 m_9 m_1^2\cr\cr
&&+\frac{3}{8} m_2 m_3^2 m_4 m_5^2 m_1+\frac{2}{3} m_2^3 m_4 m_7^2 m_1+\frac{1}{3} m_2^2 m_6 m_8^2 m_1-\frac{2}{3} m_5 m_6 m_8^2 m_1-\frac{2}{9} m_3 m_6 m_9^2 m_1\cr\cr
&&+\frac{1}{8} m_2^5 m_3^2 m_4 m_1+m_{10}^2 m_2 m_4 m_1+\frac{2}{3} m_{10} m_2^3 m_3 m_4 m_1-\frac{5}{12} m_2^3 m_3^2 m_4 m_5 m_1\cr\cr
&&-\frac{3}{2} m_{10} m_2 m_3 m_4 m_5 m_1+\frac{1}{12} m_3 m_5^3 m_6 m_1+\frac{1}{18} m_2^4 m_4^2 m_6 m_1-\frac{1}{6} m_{10} m_5^2 m_6 m_1\cr\cr
&&-\frac{1}{6} m_2^2 m_3 m_5^2 m_6 m_1-\frac{1}{6} m_{10} m_2^2 m_5 m_6 m_1+\frac{1}{6} m_2^2 m_4^2 m_5 m_6 m_1-\frac{1}{36} m_2^4 m_3 m_5 m_6 m_1\cr\cr
&&-\frac{1}{4} m_3 m_4 m_5^2 m_7 m_1-\frac{3}{2} m_{10} m_2^2 m_4 m_7 m_1-\frac{7}{12} m_2^4 m_3 m_4 m_7 m_1+\frac{1}{2} m_{10} m_4 m_5 m_7 m_1\cr\cr
&&+m_2^2 m_3 m_4 m_5 m_7 m_1-\frac{1}{36} m_2^5 m_6 m_7 m_1-\frac{1}{12} m_2 m_5^2 m_6 m_7 m_1+\frac{1}{3} m_2^3 m_5 m_6 m_7 m_1\cr\cr
&&-2 m_{10}^2 m_8 m_1-\frac{1}{6} m_2^4 m_3^2 m_8 m_1-\frac{1}{4} m_3^2 m_5^2 m_8 m_1-\frac{1}{2} m_2^2 m_7^2 m_8 m_1+\frac{1}{2} m_5 m_7^2 m_8 m_1\cr\cr
&&-m_{10} m_2^2 m_3 m_8 m_1+\frac{1}{2} m_2^2 m_3^2 m_5 m_8 m_1+\frac{3}{2} m_{10} m_3 m_5 m_8 m_1-\frac{1}{3} m_2^3 m_4 m_6 m_8 m_1\cr\cr
&&+\frac{1}{3} m_2 m_4 m_5 m_6 m_8 m_1+2 m_{10} m_2 m_7 m_8 m_1+\frac{2}{3} m_2^3 m_3 m_7 m_8 m_1-\frac{3}{2} m_2 m_3 m_5 m_7 m_8 m_1\cr\cr
&&-\frac{2}{3} m_4 m_7^2 m_9 m_1+\frac{1}{3} m_{10} m_3 m_4 m_9 m_1-\frac{1}{12} m_3^2 m_4 m_5 m_9 m_1-\frac{5}{9} m_2 m_4^2 m_6 m_9 m_1\cr\cr
&&+\frac{1}{3} m_{10} m_2 m_6 m_9 m_1+\frac{1}{18} m_2^3 m_3 m_6 m_9 m_1+\frac{5}{18} m_2 m_3 m_5 m_6 m_9 m_1+\frac{1}{3} m_2 m_3 m_4 m_7 m_9 m_1\cr\cr
&&-\frac{7}{18} m_2^2 m_6 m_7 m_9 m_1+\frac{1}{6} m_5 m_6 m_7 m_9 m_1-\frac{1}{12} m_2 m_3^2 m_8 m_9 m_1+\frac{2}{3} m_4 m_6 m_8 m_9 m_1\cr\cr
&&+\frac{1}{3} m_3 m_7 m_8 m_9 m_1+m_{10}^3+\frac{m_2^6 m_3^3}{144}+\frac{m_6 m_8^3}{3}+\frac{1}{6} m_{10} m_2^4 m_3^2-\frac{1}{16} m_2^4 m_3^2 m_4^2-\frac{1}{4} m_{10} m_2^2 m_3 m_4^2\cr\cr
&&+\frac{1}{16} m_2^2 m_3^3 m_5^2+\frac{1}{2} m_{10} m_3^2 m_5^2-\frac{1}{16} m_3^2 m_4^2 m_5^2+\frac{m_2^6 m_6^2}{216}+\frac{1}{8} m_2^2 m_5^2 m_6^2-\frac{1}{24} m_2^4 m_5 m_6^2\cr\cr
&&+\frac{3}{2} m_{10} m_2^2 m_7^2+\frac{1}{8} m_3 m_5^2 m_7^2+\frac{7}{24} m_2^4 m_3 m_7^2-\frac{1}{2} m_{10} m_5 m_7^2-\frac{1}{4} m_2^2 m_3 m_5 m_7^2-\frac{1}{8} m_2^2 m_3^2 m_8^2\cr\cr
&&-\frac{1}{2} m_{10} m_3 m_8^2+\frac{1}{8} m_3^2 m_5 m_8^2-\frac{2}{3} m_2 m_4 m_6 m_8^2+\frac{1}{2} m_2 m_3 m_7 m_8^2+\frac{m_6^2 m_9^2}{9}+\frac{3}{4} m_{10}^2 m_2^2 m_3\cr\cr
&&-\frac{1}{48} m_2^4 m_3^3 m_5-\frac{1}{2} m_{10} m_2^2 m_3^2 m_5+\frac{1}{8} m_2^2 m_3^2 m_4^2 m_5+\frac{1}{4} m_{10} m_3 m_4^2 m_5-\frac{5}{4} m_{10}^2 m_3 m_5\cr\cr
&&-\frac{1}{9} m_2^3 m_4^3 m_6-\frac{1}{12} m_2 m_3 m_4 m_5^2 m_6-\frac{1}{36} m_2^5 m_3 m_4 m_6+\frac{1}{3} m_{10} m_2 m_4 m_5 m_6+\frac{1}{6} m_2^3 m_3 m_4 m_5 m_6\cr\cr
&&-\frac{1}{12} m_2^5 m_3^2 m_7+\frac{1}{6} m_2^3 m_3 m_4^2 m_7-\frac{1}{4} m_2 m_3^2 m_5^2 m_7-2 m_{10}^2 m_2 m_7-m_{10} m_2^3 m_3 m_7\cr\cr
&&+\frac{1}{6} m_2^3 m_3^2 m_5 m_7+\frac{3}{2} m_{10} m_2 m_3 m_5 m_7+\frac{1}{12} m_4 m_5^2 m_6 m_7+\frac{1}{12} m_2^4 m_4 m_6 m_7-\frac{1}{2} m_2^2 m_4 m_5 m_6 m_7\cr\cr
&&+\frac{1}{6} m_2^3 m_3^2 m_4 m_8+\frac{1}{2} m_{10} m_2 m_3 m_4 m_8-\frac{1}{4} m_2 m_3^2 m_4 m_5 m_8-\frac{1}{6} m_{10} m_2^2 m_6 m_8+\frac{1}{2} m_2^2 m_4^2 m_6 m_8\cr\cr
&&-\frac{1}{6} m_3 m_5^2 m_6 m_8-\frac{1}{6} m_4^2 m_5 m_6 m_8+\frac{1}{2} m_{10} m_5 m_6 m_8+m_{10} m_4 m_7 m_8-\frac{1}{2} m_2^2 m_3 m_4 m_7 m_8\cr\cr
&&+\frac{1}{3} m_2 m_5 m_6 m_7 m_8-\frac{1}{18} m_2^3 m_3^3 m_9+\frac{m_7^3 m_9}{3}-\frac{1}{6} m_{10} m_2 m_3^2 m_9+\frac{1}{27} m_2^3 m_6^2 m_9-\frac{2}{9} m_2 m_5 m_6^2 m_9\cr\cr
&&-\frac{2}{3} m_2 m_3 m_7^2 m_9+\frac{1}{12} m_2 m_3^3 m_5 m_9+\frac{1}{9} m_4^3 m_6 m_9-\frac{2}{3} m_{10} m_4 m_6 m_9-\frac{2}{9} m_2^2 m_3 m_4 m_6 m_9\cr\cr
&&+\frac{1}{6} m_3 m_4 m_5 m_6 m_9+\frac{1}{3} m_2^2 m_3^2 m_7 m_9-\frac{1}{6} m_3 m_4^2 m_7 m_9+\frac{1}{2} m_{10} m_3 m_7 m_9-\frac{1}{6} m_3^2 m_5 m_7 m_9\cr\cr
&&+\frac{2}{3} m_2 m_4 m_6 m_7 m_9+\frac{1}{12} m_3^2 m_4 m_8 m_9+\frac{1}{6} m_2 m_3 m_6 m_8 m_9-\frac{2}{3} m_6 m_7 m_8 m_9-\frac{m_2^3 m_7^3}{3}\cr\cr
&&-\frac{m_3^3 m_5^3}{16}-\frac{m_5^3 m_6^2}{72}-\frac{m_3^3 m_9^2}{72}
\eea

\section{Partition Function for two matrices and $N=7$}\label{N7M2PF}

The partition function, graded by degree, for the two matrix model, at $N=7$ is given by
\bea
Z={P_{N=7,M=2}(x)\over (1-x)^2(1-x^2)^3(1-x^3)^4(1-x^4)^6(1-x^5)^8(1-x^6)^{11}(1-x^7)^8(1-x^8)^5(1-x^{10})^2(1-x^{12})}\cr\label{N7M2}
\eea
where the numerator is the following polynomial, of highest degree 180:
\bea
&&\!\!\!\!P_{N=7,M=2}(x)=1 + 3 x^6 + 12 x^7 + 22 x^8 + 44 x^9 + 73 x^{10}+ 120 x^{11}+ 
 211 x^{12}+ 360 x^{13}+ 670 x^{14}+ 1216 x^{15}\cr\cr
&&+ 2235 x^{16}+ 3984 x^{17}+7068 x^{18}+ 12240 x^{19}+ 21000 x^{20}+ 35632 x^{21}+ 59903 x^{22}+99884 x^{23}\cr\cr
&&+165235 x^{24}+ 270868 x^{25}+ 440100 x^{26}+ 708380 x^{27}+1129523 x^{28}+ 1783668 x^{29}+ 2790743 x^{30}\cr\cr
&&+ 4325256 x^{31}+6642998 x^{32}+ 10109920 x^{33}+ 15249397 x^{34}+ 22797452 x^{35}+33783642 x^{36}\cr\cr
&&+49627540 x^{37}+ 72273392 x^{38}+ 104352164 x^{39}+149391163 x^{40}+ 212071868 x^{41}\cr\cr
&&+ 298546313 x^{42}+ 416815320 x^{43}+577185131 x^{44}+ 792791176 x^{45}+ 1080209362 x^{46}\cr\cr
&&+1460138296 x^{47}+ 1958163820 x^{48}+ 2605567144 x^{49}+3440201861 x^{50}+ 4507361940 x^{51}\cr\cr
&&+5860650601 x^{52}+7562779132 x^{53}+9686250253 x^{54}+12313865320 x^{55}+15538979905 x^{56}\cr\cr
&&+19465445392 x^{57}+ 24207121182 x^{58}+29886956976 x^{59}+36635469839 x^{60}+ 44588661728 x^{61}\cr\cr
&&+53885259845 x^{62}+ 64663314196 x^{63}+ 77056113007 x^{64}+91187529388 x^{65}+ 107166805856 x^{66}\cr\cr
&&+ 125082960364 x^{67}+144998954988 x^{68}+166945826292 x^{69}+ 190916994986 x^{70}\cr\cr
&&+216863094628 x^{71}+244687455483 x^{72}+274242673996 x^{73}+ 
 305328404661 x^{74}\cr\cr
&&+ 337690723576 x^{75}+ 371023131460 x^{76}+ 404969499640 x^{77}+ 439128839913 x^{78}\cr\cr
&&+473062030700 x^{79}+506300351267 x^{80}+538355569900 x^{81}+ 568731422870 x^{82}\cr\cr
&&+596936026404 x^{83}+ 622494889513 x^{84}+ 644963930812 x^{85}+ 
 663942209186 x^{86}\cr\cr
&&+679083616940 x^{87}+ 690107291628 x^{88}+696806169316 x^{89}+ 699053377750 x^{90}\cr\cr
&&+696806169316 x^{91}+690107291628 x^{92}+ 679083616940 x^{93}+ 663942209186 x^{94}\cr\cr
&&+644963930812 x^{95}+ 622494889513 x^{96}+ 596936026404 x^{97}+ 
 568731422870 x^{98}\cr\cr
&&+538355569900 x^{99}+ 506300351267 x^{100}+473062030700 x^{101}+ 439128839913 x^{102}\cr\cr
&&+404969499640 x^{103}+371023131460 x^{104}+ 337690723576 x^{105}+ 305328404661 x^{106}\cr\cr
&&+274242673996 x^{107}+ 244687455483 x^{108}+ 216863094628 x^{109}+ 
 190916994986 x^{110}\cr\cr
&&+166945826292 x^{111}+ 144998954988 x^{112}+ 125082960364 x^{113}+ 107166805856 x^{114}\cr\cr
&&+91187529388 x^{115}+77056113007 x^{116}+ 64663314196 x^{117}+ 53885259845 x^{118}\cr\cr
&&+44588661728 x^{119}+ 36635469839 x^{120}+ 29886956976 x^{121}+ 
 24207121182 x^{122}\cr\cr
&&+19465445392 x^{123}+ 15538979905 x^{124}+12313865320 x^{125}+ 9686250253 x^{126}\cr\cr
&&+ 7562779132 x^{127}+ 
 5860650601 x^{128}+ 4507361940 x^{129}+ 3440201861 x^{130}+ 
 2605567144 x^{131}\cr\cr
&&+1958163820 x^{132}+ 1460138296 x^{133}+1080209362 x^{134}+ 792791176 x^{135}+ 577185131 x^{136}\cr\cr
&&+416815320 x^{137}+ 298546313 x^{138}+ 212071868 x^{139}+ 
 149391163 x^{140}+ 104352164 x^{141}\cr\cr
&&+ 72273392 x^{142}+49627540 x^{143}+ 33783642 x^{144}+ 22797452 x^{145}+ 15249397 x^{146}\cr\cr
&&+10109920 x^{147}+ 6642998 x^{148}+ 4325256 x^{149}+ 2790743 x^{150}+ 
 1783668 x^{151}+ 1129523 x^{152}\cr\cr
&&+ 708380 x^{153}+ 440100 x^{154}+ 
 270868 x^{155}+ 165235 x^{156}+ 99884 x^{157}+ 59903 x^{158}+ 
 35632 x^{159}\cr\cr
&&+ 21000 x^{160}+ 12240 x^{161}+ 7068 x^{162}+ 3984 x^{163}+ 
 2235 x^{164}+ 1216 x^{165}+ 670 x^{166}+ 360 x^{167}\cr\cr
&&+ 211 x^{168}+120 x^{169}+ 73 x^{170}+ 44 x^{171}+ 22 x^{172}+ 12 x^{173}+ 
 3 x^{174}+ x^{180}
\eea

\end{document}